# SST polarization model and polarimeter calibration

Jakob Selbing

June 26, 2005

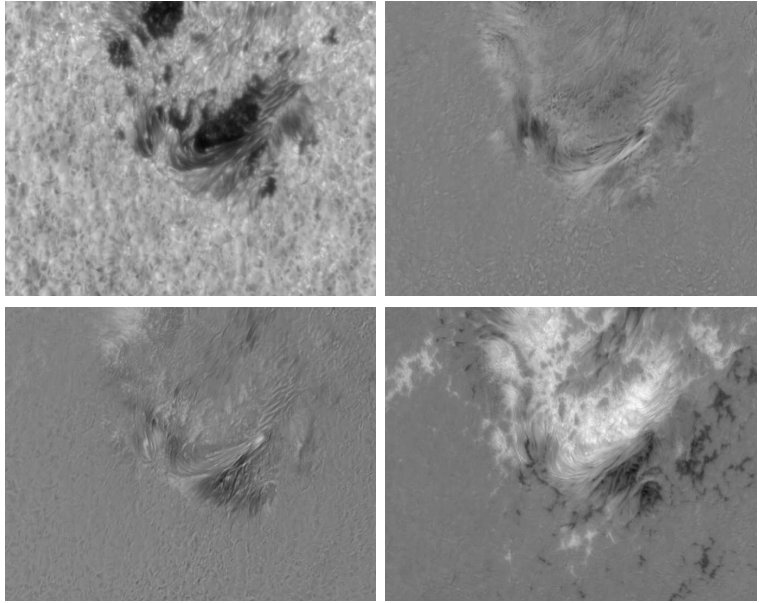

## Abstract

A telescope polarization model for the SST is developed and the parameters of this model are fitted to polarization measurements made with a 1-meter linear polarizer in front of the entrance window. In this model, the 1-meter lens is characterized by a five-parameter Müller matrix, corresponding to a retarder with arbitrary variations of the retardance and fast-axis orientation across the aperture. The resulting model is verified by measuring the telescope polarization for unpolarized input light and comparing to predictions from the polarization model. The accuracy of the prediction is within approximately 0.4% for all normalized polarization components ($Q/I$, $U/I$ and $V/I$).

The polarimeter used is based on two nematic liquid crystals and one linear polarizer, and will be used for both imaging polarimetry and spectropolarimetry. The most critical calibration is measuring the modulation matrix. This is done by inserting one linear polarizer and one rotating quarter-wave plate in the optical path before the polarimeter, and measuring the modulated intensity. The calibration of the quarter-wave plate is optimized by measuring the linear polarizer only with the polarimeter, and then minimizing the error in degree of polarization plus the residual error for the inversion of the modulation matrix by iteration of the two unknown parameters (retardance and angle offset). We find that small non-linearities in the CCD response is the major obstacle in calibrating the polarimeter.

The first full Stokes imaging polarimetry observations at the SST are shown. Comparing images before and after telescope compensation verify the telescope polarization model.

## Acknowledgements

There are mainly two persons that have been very supportive during this work: Göran Scharmer at the Institute for Solar Physics, Sweden, and Manuel Collados at the Instituto de Astrofísica de Canarias, Spain. Further, Mats Löfdahl and Peter Dettori at the Institute for Solar Physics are acknowledged for their help on technical issues, programming in ANA and writing LaTeX. For allowing publishing of their polarimetry data, Manuel Collados and Olena Khomenko (Instituto de Astrofísica de Canarias) are credited. Many thanks also to Christian Beck at the Kiepenheuer-Institut für Sonnenphysik, Germany, for advice on polarimeter calibrations.

# Contents

# Document overview

**Chapter 1** is an introduction to polarimetry in solar observations, different polarimeter designs, and problems related to polarimetry.

**Chapter 2** describes the SST polarimeter design, calibrations, alignment methods, problems with optical elements and CCD cameras.

**Chapter 3** deals with the telescope polarization model. By ignoring elements that are expected to have little influence on the polarization, a 10-parameter model is developed. These parameters are fitted to polarization data made with a linear polarizer in front of the telescope, and the model is then tested by comparing measured and predicted polarization for unpolarized light.

**Chapter 4** shows some of the first full Stokes polarimetric observations made (June 3, 2005). Images are shown both with and without telescope compensation using the telescope model, and the results indicate that the model is accurate.

**Chapter 5** is a conclusion of the work, suggestions on future improvements, and some personal comments.

**Appendix A** shows the complete telescope polarization data from May 8, 2005, together with the fitted data and residual errors.

**Appendix B** describes Stokes vector and Müller matrix formalism.

**Appendix C** gives some definitions in polarimetry.

**Appendix D** lists some of the ANA scripts used for this work.

# Chapter 1

# Introduction

Solar imaging and polarimetry is an area of research which in later years has seen a rapid improvement in high-resolution observations and with one particular telescope, the Swedish 1-meter Solar Telescope (SST), breaching the long anticipated $0''\!.1$ spatial resolution limit. Polarimetry is used primarily to measure the Sun's magnetic field, since magnetic fields induce polarization in many absorption lines. The small-scale magnetic field of the Sun gives clues about what processes control e.g. the stability and dynamics of sunspots.

In polarimetry the polarization of light is described by the Stokes vector, which is composed by four scalars: $I$, $Q$, $U$ and $V$. Stokes vector formalism is described in appendix B.

The two main areas of solar polarimetry is imaging polarimetry and spectropolarimetry. Imaging polarimetry uses modulation optics and narrowband filters to isolate wavelengths with polarization signatures. It is then possible to demodulate the images to create a 2-dimensional map of the polarization signal across the surface. Spectropolarimetry is when a spectrograph is used in conjunction with modulation optics, so that the spectrum created can be demodulated in the same way as with polarimetric imaging to retrieve the polarization components within the narrow slit of the spectrograph. The two methods are complementary: spectropolarimetry gives very detailed spectrographic and polarimetric information but only along the spectrograph slit. Imaging polarimetry retrieves such information with much lower spectral information, degraded by seeing-induced spatial cross-talk, but at sufficiently high cadence to allow evolution of small-scale magnetic fields to be followed over a large area.

The fundamental property of a polarimeter is modulation, which means converting a polarization signal into an intensity signal. The intensity is registered by a CCD camera, and using subsequent data processing one can retrieve the polarization signal. This process is called demodulation.

The modulation can be accomplished in several ways. Some solar po-larimeter designs use a rotating quarter-wave plate and a linear polarizer or polarizing beam splitter, such as the POLIS [7], or the ASP [5]. The rotation mechanism can introduce image wobbling if not carefully designed. Other polarimeters use non-rotating LCVRs[1] in front of a linear polarizer, like the LCSP [4]. The temperature stability of such LCVRs is critical, which is also an issue with the LCSP. Common are also piezo-electric modulators, as used in ZIMPOL II [3].

The major problem of solar polarimeters is seeing cross-talk, in particular from Stokes $I$ to $Q$, $U$ and $V$. Such cross-talk can be strongly reduced by using polarizing beam splitters and off-line processing to measure and compensate for seeing-induced and any other abberations.

The second biggest problem with solar polarimetry is cross-talk between the components of the Stokes vector due to inadequate knowledge about the polarization properties of the telescope and any other instrumentation, including the polarimeter itself. In [4] is a description of an iterative procedure for reducing the cross-talk in the *polarimeter*, so that a "clean" modulation of the type $I \pm Q$ etc. is obtained. Most other polarimeters, however, use a calibration procedure where the modulation is precisely measured, so that the optimum demodulation can be calculated. For example the ASP and POLIS use a linear polarizer and retarder at certain angles to produce known polarization states for which the modulation of the polarimeter is measured.

The polarization introduced by the telescope is more complicated to compensate for. First of all, for telescopes that use moving flat mirrors to deflect the beam into a stationary location, as is the case for the SST, this matrix will depend on the pointing. Since it is practically impossible to measure the matrix for sufficiently many coordinates, it is necessary to make a model of the telescope Müller matrix and try to fit parameters of the model to actual measurements. However, measuring the matrix even at one set of coordinates is not an easy task. In principle we must use optical components of the same size as the telescope aperture, and they should be rotatable. At the SST, the currently used optical element is a 1-meter diameter linear polarizer. This restricts the calibration vectors to having $V = 0$, since a linear polarizer alone cannot create circularly polarized light. This means that we can only measure the matrix partially, and will have to leave the rest for modelling.

The polarimeter described in this document uses two temperature-stabilized LCVRs and a linear polarizer. The calibration is done with a linear polarizer and a rotating quarter-wave plate, located as far up the optical path as possible to maximize the number of optical components included in the direct polarimeter calibration (instead of the telescope polarization model). The telescope polarization is compensated for using the model approach, in combination with measurements on the telescope. The polarimeter can be used for both imaging polarimetry and spectropolarimetry.

---

[1]Liquid Crystal Variable Retarders.

# Chapter 2

# Polarimetry at the SST

To use the SST [1] for spectropolarimetry and imaging polarimetry, it is necessary to calibrate the polarization properties of all components of the telescope and any instrumentation used. The calibration is divided into two parts. The first part involves calibration of the telescope, including also the Schupmann system. The second part includes the tip-tilt and adaptive mirrors, the re-imaging lens and the spectrograph or, for the imaging polarimeter, the re-imaging and filter systems. The reason for this splitting is to minimize the number of free parameters needed for modeling of the telescope polarization, the Müller matrix of which cannot be measured directly. A consequence of this choice is also that the calibration of the telescope polarization properties can be used both for the spectrograph and imaging tables, in contrast to what was the case for the SVST[2]. There also exists the possibility that the modulation voltages applied to the LCVRs[3] can be modified to at least partly compensate for the polarization properties of the instrumentation itself, in particular the tip-tilt and adaptive mirrors, making it possible to optimize the modulation efficiency.

An important priority is to perform any calibrations using solar light from the telescope and with the instrumentation (LCVRs, CCDs etc) mounted in their proper place, to the extent that is possible. Since the telescope is polarizing, this means that any calibration scheme must use as first component a linear polarizer that is either not rotated at all or, if it is rotated (between two fixed positions), the data processing is made treating the data sets obtained at different rotation angles as separate data sets (i.e. with two unknown absolute intensities). It is practical to use the orientation angle of this first polarizer as reference for the polarization measurements made, and such that the orientation angle is the same for both the spectrograph and imaging tables. However, this angle shoud also be choosen such as to make these measurements convenient with respect to e.g any polarizing beam splitters.

---

[1] Swedish 1-meter Solar Telescope.
[2] Swedish Vacuum Solar Telescope; 50 cm predecessor of Swedish 1-meter Solar Telescope.
[3] Liquid Crystal Variable Retarders.

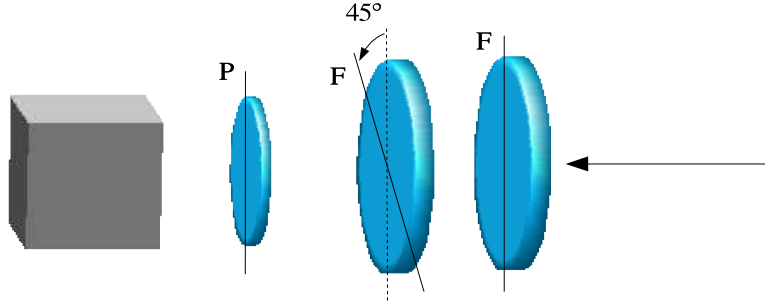

Figure 2.1: Polarimeter optical setup; light enters through two LCVRs, one linear polarizer and finally a CCD camera.

## 2.1 Polarimeter optical setup

Figure 2.1 shows the basic optical setup of the polarimeter. Two LCVRs and one linear polarizer (or polarizing beam splitter) perform the modulation of polarization into intensity. The parallel axis of the linear polarizer defines positive $Q$ for the polarimeter, and is assumed to be vertical. However, during calibration this axis will in practice be redefined by the axis of the calibrational linear polarizer. The LCVRs' fast axes are at $0°$ and $45°$ respectively. The two LCVRs and the polarizer are referred to as the modulation optics. The Müller matrix of the modulation optics is

$$\mathbf{M}_{\mathrm{mod}} = \mathbf{M}_{\mathrm{LP}} \, \mathrm{Rot}(\mathbf{M}_{\mathrm{LC2}}, \pi/4) \mathbf{M}_{\mathrm{LC1}} \qquad (2.1)$$

where $\mathbf{M}_{\mathrm{LP}}$ is the Müller matrix of a partial polarizer with extinction ratio $K_{\mathrm{LP}}$, and $\mathbf{M}_{\mathrm{LC1}}$ and $\mathbf{M}_{\mathrm{LC2}}$ are the matrices of linear retarders with retardance $\delta_{\mathrm{LC1}}$ and $\delta_{\mathrm{LC2}}$. The operation $\mathrm{Rot}(\mathbf{M}, \alpha)$ is the rotation of matrix $\mathbf{M}$ to the angle $\alpha$ (in radians). All Müller matrices used are given in Section B.3.

The CCD can only measure the $I$ (intensity) component of the light leaving the modulation optics, thus it is only the first row, $\boldsymbol{p}_{\mathrm{mod}}$, of $\mathbf{M}_{\mathrm{mod}}$ that is relevant. The intensity $I_{\mathrm{CCD}}$ registered by the CCD[4] will depend on the incident Stokes vector $\boldsymbol{S}_{\mathrm{in}}$, according to the relation

$$I_{\mathrm{CCD}} = \boldsymbol{p}_{\mathrm{mod}} \boldsymbol{S}_{\mathrm{in}} \qquad (2.2)$$

---

[4]CCD cameras can also be slightly sensitive to polarization, but this will not affect the analysis because the last element in the polarization analysis is a linear polarizer. Also their intensity response is not completely linear, which is discussed in Section 2.8.

The modulation $\boldsymbol{p}_{\text{mod}}$ is the vector

$$\boldsymbol{p}_{\text{mod}} = \begin{pmatrix} 1 & C\cos\delta_{\text{LC2}} & C\sin\delta_{\text{LC1}}\sin\delta_{\text{LC2}} & -C\cos\delta_{\text{LC1}}\sin\delta_{\text{LC2}} \end{pmatrix} \quad (2.3)$$

where

$$C = \frac{1 - K_{\text{LP}}}{1 + K_{\text{LP}}} \quad (2.4)$$

but for polarizers $K_{\text{LP}} \ll 1$, so that

$$\boldsymbol{p}_{\text{mod}} \approx \begin{pmatrix} 1 & \cos\delta_{\text{LC2}} & \sin\delta_{\text{LC1}}\sin\delta_{\text{LC2}} & -\cos\delta_{\text{LC1}}\sin\delta_{\text{LC2}} \end{pmatrix} \quad (2.5)$$

The values of $\delta_{\text{LC1}}$ and $\delta_{\text{LC2}}$ are functions of the modulation voltages $U_{\text{LC1}}$ and $U_{\text{LC2}}$ applied to the LCVRs.

To measure the Stokes vector, four measured intensities are needed. For each measurement the modulation $\boldsymbol{p}_{\text{mod}}$ is altered, by changing the modulation voltages applied to the LCVRs. By use of basic linear algebra, one can calculate the Stokes vector of the light incident on the modulation optics from the 4 measured intensities. Of course, the modulation matrix $\mathbf{P}$, which is all 4 modulations $\boldsymbol{p}_{\text{mod}}$ written row by row, must not be singular.

Note that the modulation matrix $\mathbf{P}$ and the Müller matrix $\mathbf{M}_{\text{mod}}$ of the modulation optics are two different matrices.

## 2.2  Liquid crystal retardance calibration

The modulation of the polarimeter is determined by the modulation voltages applied to the two LCVRs. In order to find the correct modulation voltages, it is necessary to find the LCVRs' voltage-to-retardance response. This is done by measuring the intensity transmitted through a setup of two polarizers with the LCVR in between, while stepping through a range of voltages (0 to 10 V, in steps of about 50 mV). It is not necessary to make a very accurate retardance calibration, since other calibrations will compensate for any errors.

Measurement are made for one LCVR at the time. The method is also adaptable to a configuration where the second polarizer is replaced by a polarizing beam splitter. The fast axis of the LCVR should be at 45° from either polarizer's axis. The intensity, as a function of voltage, is measured for both parallel and crossed polarizers. In the case of a polarizing beam splitter, one can instead make one measurement per beam, without changing the optical setup. The retardance varies with wavelength, so it is necessary to make this calibration once for each wavelength that will be used for polarimetry, using a filter in front of the CCD camera.

### 2.2.1 Calibration procedure

The procedure for aligning the LCVR and polarizers follows below. The polarizer closer to the Sun is LP1 and the polarizer (or beam splitter) closer to the camera is LP2.

1. Place LP1 and LP2 in front of the CCD (the LCVR is not yet in the beam).

2. Rotate LP1 to 0° (so that it is possible to rotate it to exactly 90° in a later stage).

3. Rotate LP2 until complete extinction occurs.

4. Rotate LP1 to 90°, making the polarizers parallel.

5. Place the LCVR between the polarizer, and rotate its fast axis to approximately 45° from the polarizers' axes (in either direction).

6. Manually adjust the driving voltage of the LCVR until minimum intensity is obtained (retardance is approx. $\pi$). This is not a critical value. The voltage should be around 1700 mV for currently used Meadowlark's #96-098 and #96-283.

7. Adjust the rotation of the LCVR until minimum intensity is obtained (fast axis is at exactly 45° from polarizers' axes).

8. Make one run of the LCVR calibration in the CCD interface, obtaining the intensity $I_\text{P}$.

9. Rotate LP1 to 0°, making the polarizers crossed.

10. Make another run of the LCVR calibration, obtaining the intensity $I_c$.

11. Block the beam and make one last calibration run, obtaining the dark level.

12. Run the LCVR calibration script `lcvr_cal.ana`. The retardance data will be saved, together with plots.

13. Repeat the whole process for the other LCVR. To save some time, it is possible to use one single dark level measurement for both LCVRs and both intensity measurements.

The intensities for the LCVR between parallel and crossed (perfect) polarizers respectively are of the form

$$I_\text{C} = I_{\text{C}0}(1 - \cos \delta_\text{LC}) \quad \text{and} \quad I_\text{P} = I_{\text{P}0}(1 + \cos \delta_\text{LC}) \qquad (2.6)$$

where $\delta_\text{LC}$ is the retardance of the LCVR.

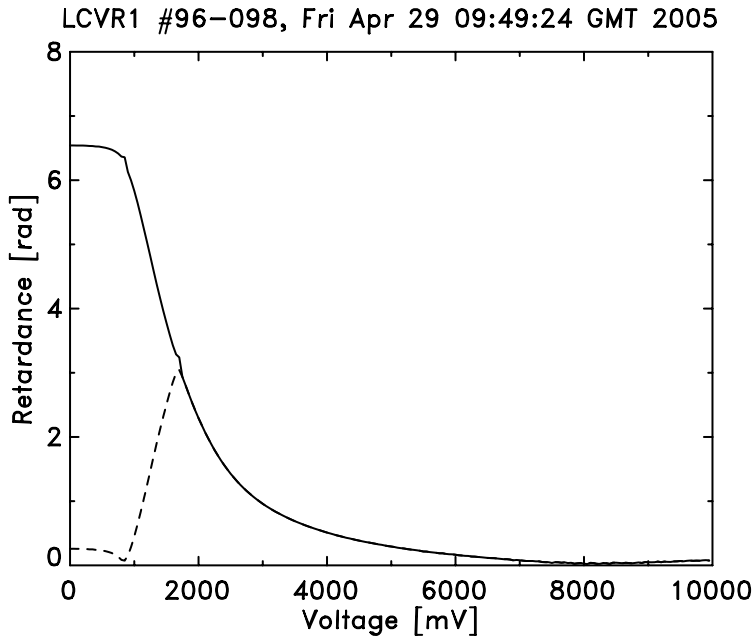

Figure 2.2: Response of LCVR #96-098.

To solve these equations for $\delta_{\mathrm{LC}}$ it is necessary to normalize both intensities, because $I_{\mathrm{C}0}$ and $I_{\mathrm{P}0}$ are not necessarily equal (due to polarization before or after polarizers). This is done by finding the two normalization factors $F_{\mathrm{C}}$ and $F_{\mathrm{P}}$, so that the sum of the two intensities is constant (e.g. 1), independently of applied voltage. This is an overdetermined linear problem, formulated as

$$F_{\mathrm{C}}\,\boldsymbol{I}_{\mathrm{C}} + F_{\mathrm{P}}\,\boldsymbol{I}_{\mathrm{P}} = \mathbf{1} \tag{2.7}$$

From the least-square solution we get

$$F_{\mathrm{C}}\boldsymbol{I}_{\mathrm{C}} = (1 - \cos\boldsymbol{\delta}_{\mathrm{LC}})/2 \quad \text{and} \quad F_{\mathrm{P}}\boldsymbol{I}_{\mathrm{P}} = (1 + \cos\boldsymbol{\delta}_{\mathrm{LC}})/2 \tag{2.8}$$

which are solved by

$$\cos\boldsymbol{\delta}_{\mathrm{LC}} = (F_{\mathrm{P}}\boldsymbol{I}_{\mathrm{P}} - F_{\mathrm{C}}\boldsymbol{I}_{\mathrm{C}}) \tag{2.9}$$

Applying arccosine to the expression above will give the retardance only when $0 \leq \delta_{\mathrm{LC}} \leq \pi$. Since the shape of the curve is predictable, software written for the calibration can "fold back" the retardance to the correct value. The ANA script `lcvr_cal.ana` uses the normalization procedure, and compensates for the retardance in a way that is most suitable for the LCVRs currently in use (Meadowlark #96-098 and #96-283). The script writes a plot to disk. One such plot is showed in figure 2.2. The dashed line shows the raw arccos value, while the solid line is the values folded back.

## 2.3  Selection of modulation voltages

When the response of the LCVRs is known, it is possible to calculate suitable modulation voltages for a given modulation matrix. This matrix, referred to as the target modulation matrix $\mathbf{T}$, cannot be arbitrarily selected, since the polarimeter setup only allows modulation of the form given in equation 2.5. In this implementation, the best resembling modulation in least-square sense is always used.

First, the modulation retardances is calculated for one modulation (one row of $\mathbf{P}$) at a time, using the least-square algorithm. Due to the periodicity of the sine and cosine functions, each state of modulation (or the closest least-square modulation) has always two equivalent solutions (not counting multiples of $2\pi$), namely

$$\begin{array}{ll} \delta_{\text{LC1}} = A & \delta_{\text{LC1}} = A + \pi \\ \delta_{\text{LC2}} = B & \text{and} \quad \delta_{\text{LC2}} = -B \end{array} \quad (2.10)$$

For a given retardance $\delta_{\text{LC}}$ we must find a modulation voltage $U_{\text{LC}}$, so that the achieved retardance is close to $\delta_{\text{LC}}$, i.e.

$$D = \delta_{\text{LC}} - \delta(U_{\text{LC}}) \approx 0 \quad (2.11)$$

where $\delta(U_{\text{LC}})$ is the retardance at voltage $U_{\text{LC}}$ measured in the LCVR retardance calibration. Multiples of $2\pi$ are ignored. In case the difference $D$ is larger than $\pi$ the value $2\pi - D$ is used as the difference instead, since in absolute terms two angles cannot be separated by more than $\pi$.

Thus, for each state of modulation we will have two sets of modulation voltages, that are equivalent in terms of modulation. To select either of the two sets, it has been suggested to compare the average modulation voltage $(U_{\text{LC1}} + U_{\text{LC2}})/2$ for each solution, because LCVRs have longer response time for large voltage amplitudes. It is thus beneficial to use the solution which has lower average modulation voltage. Another method would be to select the solution where the retardance is less sensitive to variation in modulation voltages, i.e. the derivative of the retardance with respect to the voltage is low.

The script `lcvr_set.ana` does the selection automatically, and prints suggested modulation voltages to screen (not to the polarimeter configuration file). Below is an example of the output for 6302 Å. The target matrix is

$$\begin{pmatrix} 1 & e & e & e \\ 1 & e & -e & -e \\ 1 & -e & -e & e \\ 1 & -e & e & -e \end{pmatrix} \quad (2.12)$$

where $e = 1/\sqrt{3}$, which is one of several optimum modulation matrices. The output from the script was

```
        Mode 1   Mode 2   Mode 3   Mode 4
        ------   ------   ------   ------
LC1:    2.3562   5.4978   3.9270   0.7854   [rad]
          1980     1085     1465     3370   [mV]
LC2:    0.9553   0.9553   2.1863   2.1863   [rad]
          3665     3665     2255     2255   [mV]
 or
LC1:    5.4978   2.3562   0.7854   3.9270   [rad]
          1085     1980     3370     1465   [mV]
LC2:    5.3279   5.3279   4.0969   4.0969   [rad]
          1170     1170     1495     1495   [mV]
```

where for each state of modulation both solutions are shown. The best solutions, with respect to lowest average modulation voltage, were

```
        Mode 1   Mode 2   Mode 3   Mode 4
        ------   ------   ------   ------
LC1:      1085     1980     1465     1465   [mV]
LC2:      1170     1170     2255     1495   [mV]
```

## 2.4 Polarimeter optics alignment

The alignment of the modulation optics of the polarimeter is an important, but not critical, step. All software associated with the polarimeter assumes one certain configuration of the optics, and if this differs from reality the result would be less efficient modulation, and also inability to adjust modulation voltages correctly. However the accuracy is not compromised, since the polarimeter matrix calibration directly measures the modulation, and calculates the demodulation matrix.

### 2.4.1 Alignment procedure

The only alignment that is needed is the rotation of the two LCVRs relative to the linear polarizer, LP2. The axis of LP2 is assumed to be vertical, but it can in principle be arbitrarily selected. For this purpose one extra linear polarizer (LP1) is needed in front of the LCVRs.

1. Place the extra polarizer LP1 in front of the LCVRs, and then remove the LCVRs.

2. Rotate LP1 to complete extinction of the beam.

3. Reinsert LCVR1 and rotate it so that its fast axis is approx. 45° from either polarizer's axis.

4. Adjust the driving voltage of LCVR1 so that the intensity is maximized (not critical). The retardance is around $\pi$ at this point. The voltage is around 1700 mV for currently used Meadowlark's #96-098 and #96-283.

5. Rotate LCVR1 so that its fast axis is parallel to LP2's axis, keeping the driving voltage from previous step. Fine adjust to make the intensity minimized. This makes the fast axis exactly parallel to LP2's axis.

6. Remove LCVR1, reinsert LCVR2 and rotate it to approx. 45° from either polarizer's axis.

7. Adjust the driving voltage of LCVR2 so that the intensity is maximized (not critical). The retardance is around $\pi$ at this point.

8. Rotate LP1 90° in either direction, making it parallel to LP2.

9. Adjust the rotation of LCVR2 to minimize the intensity. If the minimum is difficult to find, first adjust the driving voltage towards minimum intensity.

10. Reinsert LCVR1 and remove the extra polarizer LP1.

## 2.5 Modulation matrix calibration

The modulation matrix calibration is the most critical calibration. The goal is to precisely measure the modulation matrix of the polarimeter. This is used for offline demodulation of the four (or two, for V-only observations) modulated intensities into single Stokes vector. It can also be used for adjusting the LCVRs' modulation voltages towards more efficient modulation, although the extra polarization and reflections introduced between the calibration optics and the modulation optics of the polarimeter will need to be accounted for. Note that for magnetogram observations, the cross-talk can be measured but not completely compensated for.

To make a polarimeter matrix calibration we will use one calibrational linear polarizer (CLP) in front of one rotatable quarter-wave plate (QWP). These are placed directly after the telescope field lens (at the exit of the telescope vacuum tube) in a specially designed mount (shown in figure 2.3). These elements are referred to as the calibrational optics. The axis of the CLP will define positive $Q$ of the coordinate frame for the polarimeter modulation matrix. The QWP is mounted in a rotator stage, which is controlled by the camera PC. It is rotated in steps of 5°, and at each position the camera measures the 4 (or 6, if V-only measurements are also used) modulated intensities. It is then possible to calculate the precise modulation from the intensities, because the Stokes vector produced by the calibration

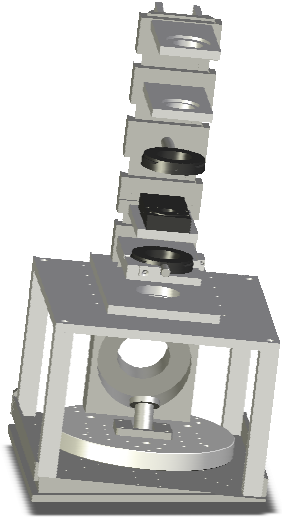

Figure 2.3: Calibration optics mount, standing on the AO system. From top: pupil re-imaging lens mount, filter mount, polarizer mount, QWP rotator stage, and 0°/90° polarizer mount. In the later configuration only the lower polarizer mount was used, above the rotator stage.

optics can be calculated theoretically, assuming their properties can be determined. The most difficult part of this calibration is to determine those properties. The extinction ratio of the CLP is measured separately, using crossed/parallel intensity measurement. Initially, a separate calibration procedure was used for measuring three parameters of the QWP: retardance, angle offset and dichroism. However, the method proved too inaccurate due to high sensitivity to CCD non-linearities; in particular the retardance could not be accurately determined even with very high SNR. Instead a different and more robust method was developed. We first assume that approximate values of the three QWP parameters are known, and that the dichroism is negligible (this is the case for the QWP currently used). The Stokes vector components produced by the calibration optics is then

$$I = I_0 \tag{2.13}$$
$$Q = I_0\, C\, (\cos^2 2\alpha_{\mathrm{QWP}} + \sin^2 2\alpha_{\mathrm{QWP}}\, \cos \delta_{\mathrm{QWP}}) \tag{2.14}$$
$$U = I_0\, C\, \cos 2\alpha_{\mathrm{QWP}}\, \sin 2\alpha_{\mathrm{QWP}}(1 - \cos \delta_{\mathrm{QWP}}) \tag{2.15}$$
$$V = I_0\, C\, \sin 2\alpha_{\mathrm{QWP}}\, \sin \delta_{\mathrm{QWP}} \tag{2.16}$$
$$\tag{2.17}$$

where
$$C = \frac{1 - K_{\mathrm{CLP}}}{1 + K_{\mathrm{CLP}}}, \tag{2.18}$$

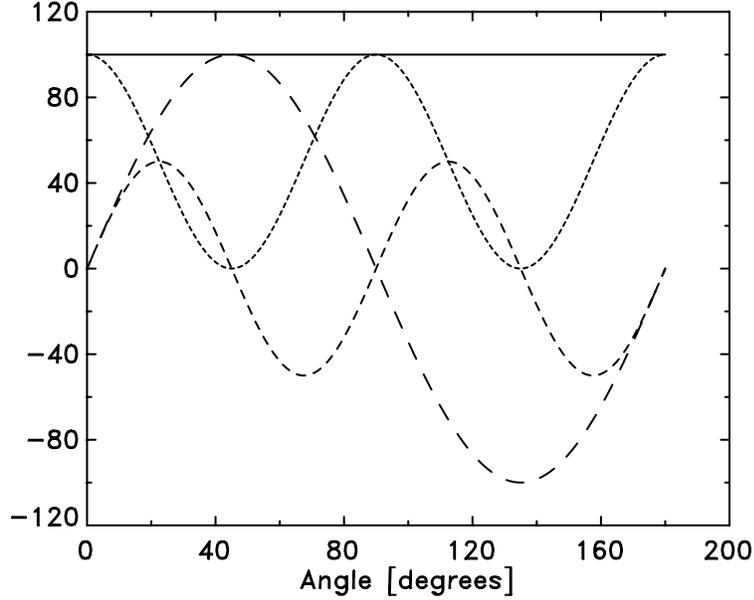

Figure 2.4: Stokes vector produced by calibrational optics with $K_{\rm LP} = 10^{-5}$ and $\delta_{\rm QWP} = \pi/2$. Solid=$I$, short-dash=$Q$, medium-dash=$U$, long-dash=$V$.

$K_{\rm CLP}$ is the extinction ratio of the CLP, and $\alpha_{\rm QWP}$ and $\delta_{\rm QWP}$ are the angle offset and the retardance of the QWP. Positive $Q$ is along the optical axis of the CLP.

Figure 2.4 shows the four Stokes components when the angle of the QWP is rotated from 0° to 180°. For each position of the QWP, one polarimetric observation is made, i.e. four intensities at different states of modulation are measured. The relation between measured intensities, modulation matrix $\mathbf{P}$, and input Stokes vector is described by

$$\begin{pmatrix} I_1 \\ I_2 \\ I_3 \\ I_4 \end{pmatrix}_n = \begin{pmatrix} P_{1,1} & P_{1,2} & P_{1,3} & P_{1,4} \\ P_{2,1} & P_{2,2} & P_{2,3} & P_{2,4} \\ P_{3,1} & P_{3,2} & P_{3,3} & P_{3,4} \\ P_{4,1} & P_{4,2} & P_{4,3} & P_{4,4} \end{pmatrix} \begin{pmatrix} I \\ Q \\ U \\ V \end{pmatrix}_n, \quad n = 1, 2, \ldots, N \quad (2.19)$$

The rows of matrix $\mathbf{P}$ are solved one by one, by treating each intensity $I_1$ to $I_4$ (or modulation 1 to 4) separately:

$$I_{k,n} = \begin{pmatrix} P_{k,1} & P_{k,2} & P_{k,3} & P_{k,4} \end{pmatrix} \begin{pmatrix} I & Q & U & V \end{pmatrix}_n^T, \quad k = 1, 2, 3, 4 \quad (2.20)$$

equivalent to

$$\begin{pmatrix} I_{k,1} \\ I_{k,2} \\ \vdots \\ I_{k,N} \end{pmatrix} = \begin{pmatrix} I_1 & Q_1 & U_1 & V_1 \\ I_2 & Q_2 & U_2 & V_2 \\ \vdots & \vdots & \vdots & \vdots \\ I_N & Q_N & U_N & V_N \end{pmatrix} \begin{pmatrix} P_{k,1} \\ P_{k,2} \\ P_{k,3} \\ P_{k,4} \end{pmatrix}, \quad k = 1, 2, 3, 4 \quad (2.21)$$

or

$$\boldsymbol{I}_k = \mathbf{M}_{\text{Stokes}} \boldsymbol{P}_k^T, \quad k = 1, 2, 3, 4 \quad (2.22)$$

where $\boldsymbol{I}_k$ is the vector of all measured values of intensity $I_k$, and $\boldsymbol{P}_k$ is row $k$ of the modulation matrix $\mathbf{P}$. This is an overdetermined linear system of equations, which has least-squares solution

$$\widetilde{\boldsymbol{P}}_k^T = (\mathbf{M}_{\text{Stokes}}^T \mathbf{M}_{\text{Stokes}})^{-1} \mathbf{M}_{\text{Stokes}}^T \boldsymbol{I}_k, \quad k = 1, 2, 3, 4 \quad (2.23)$$

One consideration, however, is the quality of the matrix $\mathbf{M}_{\text{Stokes}}$, in terms of how accurately $\boldsymbol{P}_k^T$ can be determined. This is discussed in Section 2.5.2.

The parameters of the QWP are so far only approximate, thus the measured modulation matrix will not be accurate. However, it has been determined by numerical calculation (Section 2.5.3) that

1. the residual error from the modulation matrix calculation (equation 2.23) is increased by an error in angle offset, but is unaffected by an error in retardance.

2. the error in measured degree of polarization[5] for a perfect linear polarizer producing $q = q_t$ will be approximately $\Delta_\delta (1 - q_t)$, where $\Delta_\delta$ is the error in retardance (the angle offset error is assumed to be zero).

3. the error in measured degree of polarization for a perfect linear polarizer producing $u = u_t$ will be approximately $2\Delta_\alpha u_t$, where $\Delta_\alpha$ is the error in angle offset (the retardance error is assumed to be zero).

Thus it is possible to determine the angle offset by optimizing the residual error in the modulation matrix calculation, and then determine the retardance by optimizing the degree of polarization for a linear polarizer, e.g. the CLP. Note that with a linear polarizer, $q = \cos 2\alpha$ and $u = \sin 2\alpha$, so the two errors in degree of polarization cannot cancel each other, which means that in principle only the degree of polarization could be considered and not the modulation matrix calculation.

There are several advantages with this method. First of all, determining the angle offset directly during the modulation matrix calibration is much more safe than doing two different calibrations and assuming that the angle offset is the same during both. Secondly, by only considering the *degree* of polarization, instead of the actual Stokes vector, one does not have to rely on precise angles of the CLP. Also, the performance of linear polarizers is usually very close to an ideal (partial) polarizer, with only one parameter (extinction ratio, or contrast) that can easily be determined.

---

[5]Degree of polarization $P \doteq ((Q^2 + U^2 + V^2)/I^2)^{1/2}$.

The proper algorithm for finding the QWP parameters (and modulation matrix) would then be

1. Assume approximate values of retardance (90°) and angle offset (0°) from previous measurements.

2. Minimize the residual error from the matrix calculation by iteration of angle offset.

3. Using the angle offset given in the previous step, demodulate observations of a linear polarizer, and minimize the error in degree of polarization by iterating the estimated retardance.

4. Repeat from step 2, until values do not change.

It is possible to use only the demodulated $V/I$ residual, instead of using all four modulation intensities. Another parameter that might be included in this optimization is compensation for a linear change (trend) in intensity during the matrix calibration, since polarimeter calibrations may sometimes be done when the Sun is at low elevation. Figure 2.5 shows a few of the cases. Note that both determined angle offset and retardance are the same when using only $V/I$ residual as compared to all four modulation intensities. However with the free intensity change, both parameters are slightly changed whereas $\chi^2$ is improved only for the angle offset and not for the retardance. The value $(-4.4\%)$ for the change in intensity that was obtained in this particular calibration (May 8, 15:27 UT) seems unreasonable, and therefore this parameter should not be included in the optimization.

Assuming that the optimized values are correct, the accuracy of this method is better than $\pm 0.1°$ for both angle offset and retardance. This is well within limits for this purpose.

### 2.5.1 Optimization of modulation voltages

A likely scenario is that the measured modulation is different from the target modulation. This can be due to errors in modulation voltages, misalignments of the modulation optics (LCVRs and linear polarizer), and also intrumental polarization[6] between calibrational optics and modulation optics, etc. In such a case, we can compensate for this. Note that there are two choices with this optimization:

---

[6]For the SST, the AO system is one likely source of instrumental polarization.

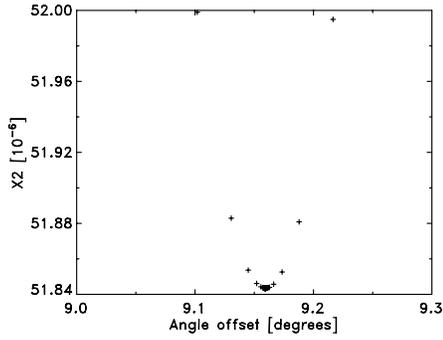
(a) Angle offset optimization; fixed intensity, full matrix used.

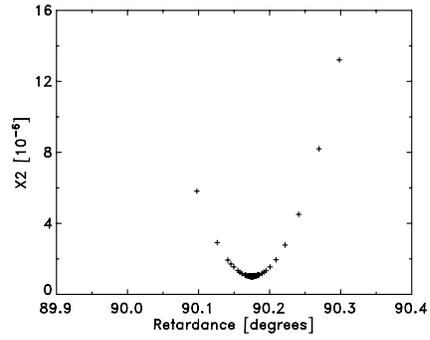
(b) Retardance optimization; fixed intensity, full matrix used.

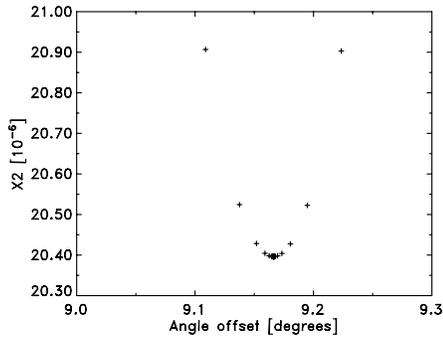
(c) Angle offset optimization; fixed intensity, only V/I residual used.

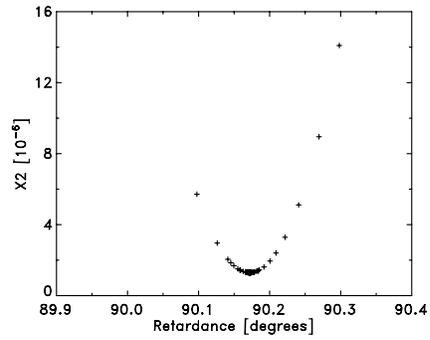
(d) Retardance optimization; fixed intensity, only V/I residual used.

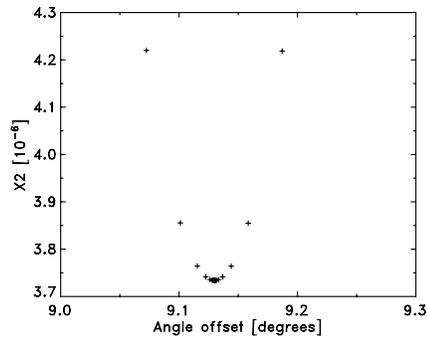
(e) Angle offset optimization; free intensity, only V/I residual used.

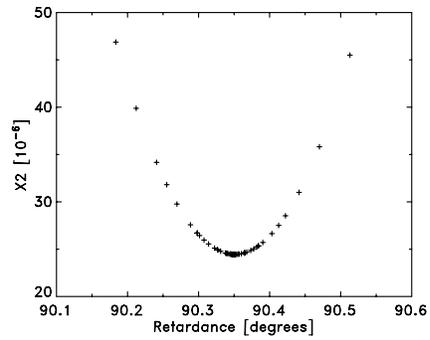
(f) Retardance optimization; free intensity, only V/I residual used.

Figure 2.5: Optimization of angle offset and retardance, in the modulation matrix calibration. Note the very distinct minima in both parameters.

1. Minimization of difference (e.g. least-square) between measured modulation and target modulation

2. Maximization of modulation efficiencies[7]

Out of these two alternatives, it is better to choose the first one (minimization of difference between measured modulation and target modulation) simply because it covers cases with both optimal and non-optimal modulation.

Due to the flexible placement of the modulation optics (either for imaging polarimetry or spectropolarimetry), there will always be the introduction of an arbitrary rotation and an arbitrary number of reflections. These are neither possible nor desirable to compensate for through the modulation voltages, because they do not change the modulation efficiencies (two states of modulation with the same efficiencies are here considered equivalent). The measured modulation will therefore be significantly different from the target modulation, but will be considered equivalent. Instead it is the non-rotational cross-talk, and possibly depolarization, that occurs in mirrors and lenses that should be eliminated. With these effects, the matrix calibration will not measure only the matrix $\mathbf{P}$ (which is what the LCVRs and linear polarizer give), but instead the product

$$\mathbf{P}_{\text{actual}} = \mathbf{P}\mathbf{P}_{\text{optics}} \tag{2.24}$$

where $\mathbf{P}_{\text{optics}}$ is the Müller matrix of the optics between the calibrational optics and the modulation optics. Note that neither $\mathbf{P}_{\text{actual}}$ nor $\mathbf{P}$ are Müller matrices, because they describe a transformation from a Stokes vector into *intensities*, not into Stokes vectors.[8] The matrix $\mathbf{P}_{\text{optics}}$ will be a combination of rotations, mirrors, beam splitters and similar elements. It is shown in appendix B.3 that any product of zero-degree mirrors $\mathbf{M}_0$ and rotations $\mathbf{R}(\alpha)$ can be only of two forms: either $\mathbf{M}_0\mathbf{R}(\alpha)$ or $\mathbf{R}(\alpha)$. Thus we can set

$$\mathbf{P}_{\text{optics}} = \mathbf{M}_0^r \mathbf{R}(\alpha) + \mathbf{M}_\epsilon \tag{2.25}$$

where $r$ is the total number of reflections modulo 2, and $\mathbf{M}_\epsilon$ is a matrix with elements $\ll 1$ and represents the unwanted cross-talk and depolarization. With this expression (2.24) becomes

$$\mathbf{P}_{\text{actual}} = \mathbf{P}(\mathbf{M}_0^r \mathbf{R}(\alpha) + \mathbf{M}_\epsilon) \tag{2.26}$$

By multiplying $\mathbf{P}_{\text{actual}}$ from right with rotations and mirrors we get

$$\begin{aligned}\mathbf{P}_{\text{actual}}\mathbf{R}(-\alpha)\mathbf{M}_0^r &= \mathbf{P}(\mathbf{M}_0^r\mathbf{R}(\alpha)\mathbf{R}(-\alpha)\mathbf{M}_0^r + \mathbf{M}_\epsilon\mathbf{R}(-\alpha)\mathbf{M}_0^r) = \ldots \\ &= \mathbf{P}(\mathbf{1} + \widetilde{\mathbf{M}}_\epsilon) \equiv \widetilde{\mathbf{P}}\end{aligned} \tag{2.27}$$

---

[7]Modulation efficiencies are defined in appendix C.

[8]Multiplying one Müller matrix $\mathbf{P}_{\text{optics}}$ with a non-Müller matrix $\mathbf{P}$ is in this case correct (but only in this particular order), because we are only replacing the expression $\mathbf{P}(\mathbf{P}_{\text{optics}}\boldsymbol{S})$ with $(\mathbf{P}\mathbf{P}_{\text{optics}})\boldsymbol{S}$.

where **1** is the unit matrix, and $\widetilde{\mathbf{M}}_\epsilon$ is still a matrix with elements $\ll 1$. In general $\alpha$ and $r$ are not known. Approximate[9] values can be found by minimizing (in least-squares sense) the difference

$$\mathbf{D} = \mathbf{P}_{\text{actual}} \mathbf{R}(-\beta) \mathbf{M}_0^j - \mathbf{T} \tag{2.28}$$

for $j \in \{0, 1\}$ and $0 \leq \beta < \pi$, where **T** is the target modulation matrix. If minimum is given by $\beta_{\min}$ and $j_{\min}$, then we have (hopefully)

$$\mathbf{P}_{\text{actual}} \mathbf{R}(-\beta_{\min}) \mathbf{M}_0^{j_{\min}} \equiv \dot{\mathbf{P}} \approx \widetilde{\mathbf{P}} \tag{2.29}$$

By adjusting the modulation voltages it might be possible to compensate for both $\widetilde{\mathbf{M}}_\epsilon$ and errors in modulation voltages, so that the difference between $\dot{\mathbf{P}}$ and **T** is reduced. Looking at each row separately, we set

$$\dot{\boldsymbol{P}}_k = \boldsymbol{T}_k + \boldsymbol{\epsilon}_k, \quad k = 1, 2, 3, 4 \tag{2.30}$$

where $\boldsymbol{\epsilon}_k$ represents an error in row $k$ of $\dot{\mathbf{P}}$. By adjusting modulation voltages we will in effect adjust the modulation retardances. We get, by linearization,

$$\dot{\boldsymbol{P}}_k^* \approx \boldsymbol{T}_k + \boldsymbol{\epsilon}_k + \Delta_{\delta_{\text{LC1,k}}} \frac{\partial \dot{\boldsymbol{P}}_k}{\partial \delta_{\text{LC1,k}}} + \Delta_{\delta_{\text{LC2,k}}} \frac{\partial \dot{\boldsymbol{P}}_k}{\partial \delta_{\text{LC2,k}}} = \boldsymbol{T}_k + \boldsymbol{C} \tag{2.31}$$

It is evident that we want $\boldsymbol{C} = 0$, i.e. we need to solve

$$\boldsymbol{\epsilon}_k + \Delta_{\delta_{\text{LC1,k}}} \frac{\partial \dot{\boldsymbol{P}}_k}{\partial \delta_{\text{LC1,k}}} + \Delta_{\delta_{\text{LC2,k}}} \frac{\partial \dot{\boldsymbol{P}}_k}{\partial \delta_{\text{LC2,k}}} = 0 \tag{2.32}$$

which is an overdetermined system of linear equation that can be solved in a least-squares sense. The modulation retardances $\delta_{\text{LC1,k}}$ and $\delta_{\text{LC2,k}}$ can be retrieved from the LCVR response data, using the modulation voltages. Since $\widetilde{\mathbf{M}}_\epsilon$ is not known, we will make the approximation

$$\frac{\partial \dot{\boldsymbol{P}}_k}{\partial \delta_{\text{LC1,k}}} \approx \frac{\partial \tilde{\boldsymbol{P}}_k}{\partial \delta_{\text{LC1,k}}} = \frac{\partial (\boldsymbol{P}_k (1 + \widetilde{\boldsymbol{M}}_{\epsilon,k}))}{\partial \delta_{\text{LC1,k}}} \approx \frac{\partial \boldsymbol{P}_k}{\partial \delta_{\text{LC1,k}}} \tag{2.33}$$

and equivalently

$$\frac{\partial \dot{\boldsymbol{P}}_k}{\partial \delta_{\text{LC2,k}}} \approx \frac{\partial \boldsymbol{P}_k}{\partial \delta_{\text{LC2,k}}} \tag{2.34}$$

so that (2.32) becomes

$$\boldsymbol{\epsilon}_k + \Delta_{\delta_{\text{LC1,k}}} \frac{\partial \boldsymbol{P}_k}{\partial \delta_{\text{LC1,k}}} + \Delta_{\delta_{\text{LC2,k}}} \frac{\partial \boldsymbol{P}_k}{\partial \delta_{\text{LC2,k}}} = 0 \tag{2.35}$$

---

[9] Most likely $r$ will be precisely determined, and $\alpha$ approximate.

When the retardance adjustments have been calculated, the corresponding voltage adjustments can be calculated by linear approximation of the LCVRs' responses, so that

$$\Delta_{U_{\text{LC1,k}}} = \Delta_{\delta_{\text{LC1,k}}} \left(\frac{\mathrm{d}\delta_{\text{LC1}}}{\mathrm{d}U_{\text{LC1}}}\right)^{-1} \quad (2.36)$$

and

$$\Delta_{U_{\text{LC2,k}}} = \Delta_{\delta_{\text{LC2,k}}} \left(\frac{\mathrm{d}\delta_{\text{LC2}}}{\mathrm{d}U_{\text{LC2}}}\right)^{-1} \quad (2.37)$$

To conclude, the whole process is

1. Measure the modulation matrix $\mathbf{P}_{\text{actual}}$.

2. Calculate $\dot{\mathbf{P}}$, which is an approximation of $\mathbf{P}(\mathbf{1} + \widetilde{\mathbf{M}_\epsilon})$.

3. Calculate the modulation errors $\boldsymbol{\epsilon}_k = \dot{\boldsymbol{P}}_k - \boldsymbol{T}_k$ for $k = 1, 2, 3, 4$.

4. Estimate modulation retardances $\delta_{\text{LC1,k}}$ and $\delta_{\text{LC2,k}}$ for $k = 1, 2, 3, 4$, using the LCVR response data and current modulation voltages.

5. Solve (2.35) for $\Delta_{\delta_{\text{LC1,k}}}$ and $\Delta_{\delta_{\text{LC2,k}}}$ for $k = 1, 2, 3, 4$.

6. Estimate voltage adjustments using linear approximation of the LCVRs' response.

7. Apply voltage corrections and make a new matrix calibration.

The whole process could be repeated several times, but since one matrix calibration takes rather long time (approximately 20 minutes), more than one iteration is probably not desired.

### 2.5.2 Orthogonality of calibrational Stokes vectors

One demand imposed during the matrix calibration is that the components of the calibrational Stokes vectors are sufficiently independent. The condition number (ratio between largest and smallest singular value) of the matrix $\mathbf{M}_{\text{Stokes}}$ formed by all Stokes vectors is a good measure of orthogonality. Low condition number means better properties for inversion (low error sensitivity). The setup only allows one free variable: the angle of the QWP. The angle will be stepped through at some resolution, and up to a certain maximum angle. Table 2.1 shows the condition number for a few combinations of step size and maximum angle. The conclusion is that one can use any step from 5° to 35°, with a maximum angle of at least 180°, with approximately the same accuracy.

|            | Range   |       |       |       |
| **Resolution** | 90°     | 180°  | 270°  | 360°  |
| --- | --- | --- | --- | --- |
| 5°         | 18.43   | 3.65  | 3.82  | 3.65  |
| 10°        | 16.39   | 3.68  | 3.83  | 3.65  |
| 15°        | 15.11   | 3.71  | 3.85  | 3.66  |
| 20°        | 38.41   | 3.74  | 3.80  | 3.68  |
| 25°        | 110.87  | 3.70  | 3.80  | 3.62  |
| 30°        | 16.23   | 3.83  | 3.92  | 3.71  |
| 35°        | 4.77†   | 3.87  | 3.83  | 3.58  |
| 40°        | 9.10†   | 5.24  | 4.19  | 3.74  |

Table 2.1: Condition number for calibrational Stokes vector matrix, at different resolution and range of QWP rotation. † indicates underdetermined matrix.

### 2.5.3 Error analysis for modulation matrix

The critical part of the matrix calibration is the value of retardance and angle offset of the QWP (the dichroism is assumed to be negligible). These parameters are determined in the matrix calibration. The influence of errors in the determined parameters has been examined numerically, by assuming a certain modulation, applying it to a set of calibrational Stokes vectors to retrieve modulated intensities, assuming an error in retardance or angle offset, calculating the modulation (which will be slightly wrong), and multiplying its inverse with the true modulation to retrieve a cross-talk matrix. The analysis appears valid for *any* modulation, because the cross-talk matrix does not change when the true modulation is changed. The QWP is assumed to have a true retardance of $\pi/2$ and angle offset 0, and calibrational Stokes vectors are produced at positions 0°, 5°, ..., 360° of the QWP.

If the true modulation (one modulation state) is

$$\boldsymbol{p}_{\mathrm{mod}} = \begin{pmatrix} p_I & p_Q & p_U & p_V \end{pmatrix} \tag{2.38}$$

then, with an error $\Delta_\delta$ (in radians) in **retardance**, the measured modulation will be

$$\tilde{\boldsymbol{p}}_{\mathrm{mod}} = \begin{pmatrix} \tilde{p}_I & \tilde{p}_Q & \tilde{p}_U & \tilde{p}_V \end{pmatrix} \tag{2.39}$$

with

$$\tilde{p}_I \approx p_I + \Delta_\delta\, p_Q \tag{2.40}$$
$$\tilde{p}_Q \approx p_Q\,(1 - \Delta_\delta) \tag{2.41}$$
$$\tilde{p}_U \approx p_U\,(1 - \Delta_\delta) \tag{2.42}$$
$$\tilde{p}_V \approx p_V \tag{2.43}$$

but the error $\Delta_\delta$ does *not* cause larger residual errors in the modulation matrix calculation. Linearization of the product between the inverted mea-sured modulation matrix and the true modulation matrix gives the following relation between measured and true Stokes components:

$$I_m \approx I_t \tag{2.44}$$
$$Q_m \approx Q_t \left(1 + \Delta_\delta\right) - \Delta_\delta I_t \tag{2.45}$$
$$U_m \approx U_t \left(1 + \Delta_\delta\right) \tag{2.46}$$
$$V_m \approx V_t \tag{2.47}$$

For normalized Stokes vector, the relation is

$$q_m \approx q_t \left(1 + \Delta_\delta\right) - \Delta_\delta \tag{2.48}$$
$$u_m \approx u_t \left(1 + \Delta_\delta\right) \tag{2.49}$$
$$v_m \approx v_t \tag{2.50}$$

The measured degree of polarization will be approximately

$$P_m \approx P_t + \Delta_\delta \frac{q_t \left(q_t - 1\right) + u_t^2}{P_t} \tag{2.51}$$

In the case of an error $\Delta_\alpha$ (in radians) in the **angle offset**, the situation is different. The error *does* cause a larger residual error in the modulation matrix calculation. The Stokes components are now

$$I_m \approx I_t \tag{2.52}$$
$$Q_m \approx Q_t - 4\,\Delta_\alpha\,U_t \tag{2.53}$$
$$U_m \approx U_t - 2\,\Delta_\alpha\,I_t + 4\,\Delta_\alpha\,Q_t \tag{2.54}$$
$$V_m \approx C_{\text{IV}}\,\Delta_\alpha\,I + C_{\text{QV}}\,\Delta_\alpha\,Q_t + V_t \tag{2.55}$$

For normalized Stokes vector, the relation is

$$q_m \approx q_t - 4\,\Delta_\alpha\,u_t \tag{2.56}$$
$$u_m \approx u_t - 2\,\Delta_\alpha + 4\,\Delta_\alpha\,q_t \tag{2.57}$$
$$v_m \approx C_{\text{IV}}\,\Delta_\alpha + C_{\text{QV}}\,\Delta_\alpha\,q_t + v_t \tag{2.58}$$

The constants $C_{\text{IV}}$ and $C_{\text{QV}}$ are non-integer, and depend on e.g. the QWP angle resolution. Using the original 5° resolution gives

$$C_{\text{IV}} = -0.02685 \tag{2.59}$$
$$C_{\text{QV}} = \phantom{-}0.1068 \tag{2.60}$$

but with 10° the values are

$$C_{\text{IV}} = -0.05150 \tag{2.61}$$
$$C_{\text{QV}} = \phantom{-}0.2049 \tag{2.62}$$

Thus larger angle resolution causes higher sensitivity to errors in angle offset for the $V$ component, but in practice the effect is negligible (sensitivity is much higher for $Q$ and $U$).

The measured degree of polarization will be approximately

$$P_m \approx P_t + \Delta_\alpha \frac{-4\,u_t + v_t\,(C_{\text{IV}} + C_{\text{QV}}\,q_t)}{P_t} \tag{2.63}$$

Several interesting observations are made. One is that the accuracy of the measured $V$ component is insensitive to small errors in retardance, and only weakly sensitive to errors in angle offset. Thus it is mainly $I$, $Q$ and $U$ that are affected. Also, the error in angle offset affects the residual in the modulation matrix calculation. It is therefore possible to correct for this error by searching for an angle offset that gives lower residual errors.

For the retardance error, it is possible to exploit the almost complete polarization of the linear polarizers. If we measure the Stokes vector for the CLP at an angle $\alpha$, we can assume that $V = 0$ and that the true degree of polarization $P_t$ is 1, i.e. $P_t = q_t^2 + u_t^2 = 1$. However with errors in retardance, we will instead measure

$$P_m = P_t + \Delta_\delta \frac{(q_t(q_t - 1) + u_t^2)}{P_t} = 1 + \Delta_\delta\,(1 - q_t) \tag{2.64}$$

Note that if $q_t = 1$ the degree of polarization is always 1, even with an error in retardance. When $q_t = -1$ instead, we will measure a degree of polarization of $1 + 2\,\Delta_\delta$.

## 2.6 Polarizer imperfections

During testing of calibrations, several potential problems were discovered. When the two Meadowlark polarizers were examined visually, a great variation of extinction could be seen across the aperture, depending on both the order of the specific polarizers and front side used. The polarizers have text and axis indication on one side, and it had been assumed that this was the entrance side. It will be referred to as the front side, and the other side simply the back side. The orientation where light enters the back side will be called "flipped" orientation.

To completely test the performance (extinction ratio and homogenity), there are 8 cases: 2 possible orders, and 2 ways of turning each polarizer. Table 2.2 shows the performance for all cases, by visual inspection. The conclusion is that the configuration in case F should always be used, i.e. polarizer #2 should always be first, and #1 second and flipped.

| Case | First polarizer | Second polarizer | Performance |
|------|-----------------|------------------|-------------|
| A    | #1              | #2               | bad         |
| B    | #1              | #2, flipped      | good        |
| C    | #1, flipped     | #2               | very bad    |
| D    | #1, flipped     | #2, flipped      | very bad    |
| E    | #2              | #1               | very bad    |
| F    | #2              | #1, flipped      | very good   |
| G    | #2, flipped     | #1               | very bad    |
| H    | #2, flipped     | #1, flipped      | bad         |

Table 2.2: Performance of Meadowlark polarizers for all possible configurations.

## 2.7 Quarter-wave plate imperfections

When the QWP was visually examined by rotating it between crossed polarizers, the optical axis appeared to vary slightly across its aperture. This was seen as a sweeping movement of the area of complete extinction. It was first believed that this effect caused some of the problems. To compensate for this, a new model of a retarder was suggested, in which the Müller matrix is the average of many retarders with a slight variation of orientation of the optical axis. The effect of such a retarder, compared to the ideal case, is a slight depolarization of $Q$, and reduced cross-talk between $U$ and $V$. However this model did not improve the accuracy of the model fitting, and therefore it is believed that the effects are not significant, at least not for this retarder.

## 2.8 Compensation of CCD non-linearities

During testing of the calibration methods, it was discovered that the CCD cameras seemed to have significantly non-linear response (NLR). At first, camera XIII was used, and during those calibrations the retardance was around 1.589–1.596 radians (91.04°–91.44°). After problems with the calibration, camera XII was tested instead. For this camera the retardance was estimated at 1.619 radians (92.8°) when the signal maximum was around 160 counts, and 1.606 radians (92.0°) for a signal of 320 counts.[10] Other indications of NLR were seen when measuring intensity for rotating QWP between crossed polarizers, but *not* for parallel polarizers (this was first used as a part of the calibrations, to measure the retardance and angle of the QWP). These effects were difficult to explain in any other way than with NLR of the cameras.

These non-linearities are most likely not a *major* problem for solar obser-vations, because the intensity range is rather limited. However, even though the response of the CCD may be considered perfectly linear over the range of intensities used, and offset (bias) may still be needed to characterize this response. During calibrations the intensity varies between in principle zero and up to ca. 80% of saturation, which can make the effects significant.

---
[10] Both cameras have 10-bit resolution.

To compensate for the non-linearity two methods are suggested; the double exposure method and the direct measurement method. Both are described here. *However it should be stressed that the results from tests were highly inconsistent, both when measuring and compensating non-linearity.* It could be that camera XIII, which was primarily used, has very small non-linearities. *Also, it has been concluded that the direct measurement method is not useful.* Instead the double exposure method is believed to be the only working method, however it needs a carefully designed implementation.

### 2.8.1 Double exposure method

With this method [2], we need two measurements of the intensity, at different exposures or, alternatively, using different ND filters. In this discussion the response model is an n:th order polynomial, but the method is easily adaptable to any response model.

Thus the measured intensity $I_\mathrm{m}$ is modelled as

$$I_\mathrm{m} = k\, I_\mathrm{t} + c_2\, k^2\, I_t^2 + \ldots + c_n\, k^n\, I_\mathrm{t}^n = f(k, I_\mathrm{t}, \mathbf{c}) \qquad (2.65)$$

where $I_t$ is true intensity, $k$ is a proportionality constant, and the vector $\mathbf{c} = \begin{pmatrix} c_2 & c_3 & \ldots & c_n \end{pmatrix}$. We now want to find all coefficients $\mathbf{c}$. Assuming a current estimation of all parameters, indicated as $\tilde{k}$, $\tilde{I}_\mathrm{t}$ etc., the currently replicated measured intensity, $\tilde{I}_\mathrm{m}$, would be

$$\tilde{I}_\mathrm{m} = f(\tilde{k}, \tilde{I}_\mathrm{t}, \tilde{c}_i) \qquad (2.66)$$

We define the error function $\epsilon_\mathrm{m}$ as

$$\epsilon_\mathrm{m} = I_\mathrm{m} - \tilde{I}_\mathrm{m} \qquad (2.67)$$

Now we want to change the current estimation of each parameter by a value $\Delta\tilde{k}$, $\Delta\tilde{I}_\mathrm{m}$ and $\Delta\tilde{c}_i$ so that the error function is reduced to zero. Linearization of the error function gives

$$\epsilon_\mathrm{m} = I_\mathrm{m} - (\tilde{I}_\mathrm{m} + \Delta\tilde{k}\,\frac{\partial \tilde{I}_\mathrm{m}}{\partial \tilde{k}} + \Delta\tilde{I}_\mathrm{t}\,\frac{\partial \tilde{I}_\mathrm{m}}{\partial \tilde{I}_\mathrm{t}} + \Delta\tilde{c}_2\,\frac{\partial \tilde{I}_\mathrm{m}}{\partial \tilde{c}_2} + \ldots + \Delta\tilde{c}_n\,\frac{\partial \tilde{I}_\mathrm{m}}{\partial \tilde{c}_2}) \qquad (2.68)$$

Setting $\epsilon_\mathrm{m} = 0$ and re-arranging gives the equation

$$I_\mathrm{m} - \tilde{I}_\mathrm{m} = \Delta\tilde{k}\,\frac{\partial \tilde{I}_\mathrm{m}}{\partial \tilde{k}} + \Delta\tilde{I}_\mathrm{t}\,\frac{\partial \tilde{I}_\mathrm{m}}{\partial \tilde{I}_\mathrm{t}} + \Delta\tilde{c}_2\,\frac{\partial \tilde{I}_\mathrm{m}}{\partial \tilde{c}_2} + \ldots + \Delta\tilde{c}_n\,\frac{\partial \tilde{I}_\mathrm{m}}{\partial \tilde{c}_2} \qquad (2.69)$$

This linear equation is, obviously, underdetermined since we have $n+1$ unknowns but only one equation. We could make more measurements, but each time we either measure the same thing, in which case we only duplicate equation 2.69, or measure a different $I_t$, in which case the number of unknowns is increased. *The solution is to make M measurements, each time for a different true intensity $I_t$, then change the exposure time (or intensity) and make the exact same M measurements.* This way there is only *one* additional unknown in the second measurement, namely the relative change in exposure, while the number of equations is increased with $M$. Thus we get $2M$ equations, and $n+2$ unknowns, which allows for determining an arbitrary number of coefficients in the response model as long as the number of measurements (at each exposure) is large enough.

If we write the two measurements as vectors we get the following system of *non-linear* equations

$$\boldsymbol{I}_{m,\,1} = f(k, \boldsymbol{I}_t, \boldsymbol{c}) \tag{2.70}$$
$$\boldsymbol{I}_{m,\,2} = f(k_r\, k, \boldsymbol{I}_t, \boldsymbol{c}) \tag{2.71}$$

However, the factor $k$ is redundant so we can assume $k=1$, i.e.

$$\boldsymbol{I}_{m,\,1} = f(1, \boldsymbol{I}_t, \boldsymbol{c}) \tag{2.72}$$
$$\boldsymbol{I}_{m,\,2} = f(k_r, \boldsymbol{I}_t, \boldsymbol{c}) \tag{2.73}$$

Now the linear system of equations is

$$\boldsymbol{I}_{m,\,1} - \tilde{\boldsymbol{I}}_{m,\,1} = \Delta \tilde{\boldsymbol{I}}_t \frac{\partial \tilde{\boldsymbol{I}}_{m,\,1}}{\partial \tilde{\boldsymbol{I}}_t} + \Delta \tilde{c}_2 \frac{\partial \tilde{\boldsymbol{I}}_{m,\,1}}{\partial \tilde{c}_2} + \ldots + \Delta \tilde{c}_n \frac{\partial \tilde{\boldsymbol{I}}_{m,\,1}}{\partial \tilde{c}_2} \tag{2.74}$$

$$\boldsymbol{I}_{m,\,2} - \tilde{\boldsymbol{I}}_{m,\,2} = \Delta \tilde{k}_r \frac{\partial \tilde{\boldsymbol{I}}_{m,\,2}}{\partial \tilde{k}_r} + \Delta \tilde{\boldsymbol{I}}_t \frac{\partial \tilde{\boldsymbol{I}}_{m,\,2}}{\partial \tilde{\boldsymbol{I}}_t} + \Delta \tilde{c}_2 \frac{\partial \tilde{\boldsymbol{I}}_{m,\,2}}{\partial \tilde{c}_2} + \ldots + \Delta \tilde{c}_n \frac{\partial \tilde{\boldsymbol{I}}_{m,\,2}}{\partial \tilde{c}_2} \tag{2.75}$$

Since the linearization is an approximation, solving the system will not give exactly zero in the error functions. Instead the system is iterated to convergence. The coefficients $\boldsymbol{c}$ in the response model is the primary result. Once these are found it should be easy to construct a simple operation for correcting the non-linear response, e.g. approximating the solving of the n:th order polynomial equation with another polynomial function.

The elements of $\boldsymbol{I}_{m,1}$ and $\boldsymbol{I}_{m,2}$ could be averages over an evenly illuminated area of the CCD while varying the intensity by e.g. rotating polarizers, or averages of binned pixels in two images with *inhomogenous* illumination. In both cases the signal is changed either by using different exposure times or using different ND filters.

For the polynomial model, the derivatives are

$$\frac{\partial \tilde{\boldsymbol{I}}_{\mathrm{m},\,1}}{\partial \tilde{\boldsymbol{I}}_{\mathrm{t}}} = 1 + 2\,\tilde{c}_2\,\tilde{\boldsymbol{I}}_{\mathrm{t}} + 3\,\tilde{c}_3\,\tilde{\boldsymbol{I}}_{\mathrm{t}}^{\,2} + 4\,\tilde{c}_4\,\tilde{\boldsymbol{I}}_{\mathrm{t}}^{\,3} + \ldots + n\,\tilde{c}_n\,\tilde{\boldsymbol{I}}_{\mathrm{t}}^{\,n-1} \qquad (2.76)$$

$$\frac{\partial \tilde{\boldsymbol{I}}_{\mathrm{m},\,1}}{\partial \tilde{c}_i} = \tilde{\boldsymbol{I}}_{\mathrm{t}}^{\,i} \qquad (2.77)$$

$$\frac{\partial \tilde{\boldsymbol{I}}_{\mathrm{m},\,2}}{\partial \tilde{k}_{\mathrm{r}}} = \tilde{\boldsymbol{I}}_{\mathrm{t}} + 2\,\tilde{c}_2\,\tilde{k}_{\mathrm{r}}\tilde{\boldsymbol{I}}_{\mathrm{t}}^{\,2} + 3\,\tilde{c}_3\,\tilde{k}_{\mathrm{r}}^{2}\tilde{\boldsymbol{I}}_{\mathrm{t}}^{\,3} + 4\,\tilde{c}_4\,\tilde{k}_{\mathrm{r}}^{3}\tilde{\boldsymbol{I}}_{\mathrm{t}}^{\,4} + \ldots + n\,\tilde{c}_n\,\tilde{k}_{\mathrm{r}}^{n-1}\tilde{\boldsymbol{I}}_{\mathrm{t}}^{\,n} \qquad (2.78)$$

$$\frac{\partial \tilde{\boldsymbol{I}}_{\mathrm{m},\,2}}{\partial \tilde{\boldsymbol{I}}_{\mathrm{t}}} = \tilde{k}_{\mathrm{r}} + 2\,\tilde{c}_2\,\tilde{k}_{\mathrm{r}}^{2}\,\tilde{\boldsymbol{I}}_{\mathrm{t}} + 3\,\tilde{c}_3\,\tilde{k}_{\mathrm{r}}^{3}\,\tilde{\boldsymbol{I}}_{\mathrm{t}}^{\,2} + 4\,\tilde{c}_4\,\tilde{k}_{\mathrm{r}}^{4}\,\tilde{\boldsymbol{I}}_{\mathrm{t}}^{\,3} + \ldots + n\,\tilde{c}_n\,\tilde{k}_{\mathrm{r}}^{n}\,\tilde{\boldsymbol{I}}_{\mathrm{t}}^{\,n-1} \qquad (2.79)$$

$$\frac{\partial \tilde{\boldsymbol{I}}_{\mathrm{m},\,2}}{\partial \tilde{c}_i} = \tilde{k}_{\mathrm{r}}^{i}\,\tilde{\boldsymbol{I}}_{\mathrm{t}}^{\,i} \qquad (2.80)$$

Note that the vector notation in the equations should be interpreted as "for each element in".

This method has, however, not yet been tested properly.

### 2.8.2 Direct measurement method

This method did not work when tested, but it is explained here anyway, since it is considered relevant for this report. In this method one rotating linear polarizer between two fixed parallel polarizers creates a modulation of the intensity which is very well-defined, since good linear polarizers perform very close to an ideal partial polarizer.

Consider model fitting to the measured intensity, *without* any modelling of non-linear response. There are in principle only two unknown parameters: the absolute intensity and the angle offset of the rotated polarizer. The fixed polarizers are assumed to be exactly parallel, and the extinction ratios of the polarizers are assumed to be known. If the actual response of the camera is slightly non-linear the fitting will not be perfect, however the angle offset will be accurately determined because the non-linearity is presumably small and introduces no asymmetry that could be interpreted as angle offset. The absolute intensity will simply be the one that gives the smallest $\chi^2$ value.

The residual, which is data minus the replicated data, will (in theory) represent the non-linearity error in the camera compared to *some* linear response. But if this residual can be described as a function of measured signal, then one can easily correct any measured signal. The correction that will be added to the measured signal is simply the error for that signal with a change of sign. In practice it is easier to fit some function to the error – it turned out that a 5th order polynomial is enough.

This calibration was carried out for camera XIII, at 0 dB gain and at wavelength 6302 Å. Figure 2.6(a) shows a plot of the data, the fitted model

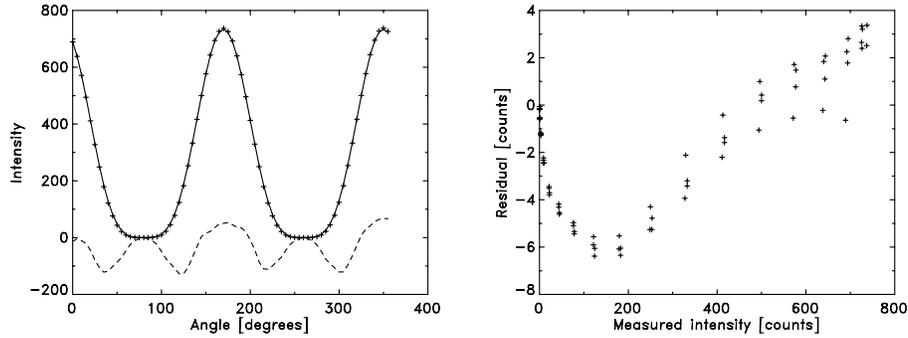

(a) Intensity for one rotating polarizer between parallel polarizers; crosses=original data, solid=fitted model, dashed=residual times 20.

(b) Residual from fit to intensity for rotating polarizer between parallel polarizer, as function of measured intensity.

Figure 2.6: Results of CCD non-linearity characterization using the direct measurement method. Note that this method has proved inaccurate, and that these plots most likely do not show the actual non-linearity.

and the residual multiplied by 20 (for visibility). The systematic behavior of the residual indicates NLR[11]. Figure 2.6(b) shows the residual, as a function of measured intensity. Note that a higher response than 100% might seem unreasonable, but this is only because the "correct" linear response is, in this case, selected to give a least-squares solution in the algorithm. It could also be chosen so that the response is always lower than 100%.

This method has one advantage: the response does not have to be modelled. However, testing of this method indicated that it did *not* measure the NLR correctly. It has not been determined exactly what causes the residual, but it is possible that a higher-quality rotating linear polarizer is needed (there were only low-quality polarizers available). Another problem is that the polarizers have to be exactly parallel. Any misalignment will cause an asymmetric residual (i.e. not systematic as a function of input signal). This, however, is not the case here. Also, errors in the rotation of the polarizer will be interpreted as non-linearity.

The conclusion is that this method cannot be used. Instead the double exposure method is considered the only alternative.

## 2.9 Characterization of linear polarizers

Some characterizations of linear polarizers were made. The angle between the optical axis and the indicated axis of all linear polarizers should be measured. The angle is critical for two polarizers: the large 1-meter polarizer and the calibrational polarizer. Also, it is useful to measure the extinction ratios of the polarizer.

---

[11]Other possible sources are e.g. polarizer imperfections.

To measure the angle, a polarizing beam splitter (PBS) was used. The direction along which the PBS divides the image is exactly parallel or orthogonal to the polarization axes of either beam. The idea of the method is to align the polarizer so that its indicated axis is parallel to the direction of the split. For this purpose a thin metal rod was taped to the polarizer. The rod could be centered in the axis indents very precisely. Then the angle of minimum intensity relative to the rod was examined. Since it was difficult to find the exact minimum, two positions (one on each side of the minimum) of equal intensity was searched for instead. It was also important to measure in the center of rotation, because the polarizers are not completely homogenous across their surface.

It turns out that for the two Meadowlark polarizers, the angle was $+5.43°$ for polarizer #1 (which was surprisingly much) and $+0.133°$ for polarizer #2. Both these values are average of several independent measurements. The accuracy is approx. $±0.07°$. Positive angle means optical axis is counterclockwise from indicated axis, when polarizer is seen from the front (the side with the axis markings and the label "Meadowlark Optics"). The software needs to include the de-rotation as a step in the demodulation process. Note that with this sign convention, the de-rotation is done using the same angle, i.e. for a misalignment of $+0.133°$ the Stokes vector is measured in a coordinate frame that is rotated $-0.133°$ as seen by the polarimeter, thus the de-rotation is made with the angle $+0.133°$.

For the 1-meter polarizer, the Meadowlark #2 polarizer was used as reference. The large polarizer is made out of two pieces of sheet polarizers. These sheet polarizers are claimed to have the optical axis exactly at $90°$ from the edge of the sheet. A measurement of the optical axis revealed that this angle was in principle perfect, with no measurable error ($< 0.2°$).

The extinciton ratio of the Meadowlark polarizers was measured at 630 nm, simply by measuring intensity $I_\text{P}$ and $I_\text{C}$ with parallel and crossed axes respectively. The ratio $C$ is then

$$C = \frac{I_\text{C}}{I_\text{P}} = \frac{K_\text{LP1} + K_\text{LP2}}{1 + K_\text{LP1} K_\text{LP2}} \approx K_\text{LP1} + K_\text{LP2} \tag{2.81}$$

Different exposures were needed because of the large dynamic range. Measurements gave $C = 4.94 \cdot 10^{-5}$, and assuming that the polarizers have the same extinction ratio, we get

$$K_\text{LP1} = K_\text{LP2} = 2.47 \cdot 10^{-5} \tag{2.82}$$

For measurement on the 1-meter polarizer, one of the Meadowlark polarizers was used, giving a value of the *local* extinction ration of

$$K_\text{1MP} = 9.06 \cdot 10^{-5} \tag{2.83}$$

# Chapter 3

# Telescope polarization at the SST

## 3.1 Telescope modelling

### 3.1.1 General telescope model

The telescope polarization model includes all optics that cannot be accounted for in the polarimeter matrix calibration. The AO system – tip/tilt-mirror and deformable mirror – and the re-imaging lens are *not* included since they are included in the matrix calibration simply by placing the calibration optics *before* these elements. The model includes the main lens, elevation mirror, azimuth mirror, field mirror, Schupmann corrector, and field lens. To model the telescope we will use the coordinate frame which is fixed to the main lens, since it is in this frame that we will produce calibration vectors (since the polarizer will be attached in front of the lens). Positive $Q$ for the telescope matrix is vertical when the turret is set to $0°$ elevation, and fixed to the lens. After demodulation and telescope compensation, the positive $Q$ in the images will be along this direction.

For convenience, Müller matrices for mirrors will be denoted $\mathbf{M}$ and a subscript, and for lenses we will instead use $\mathbf{L}$ and a subscript.

It can be useful to change coordinate frame so that positive $Q$ is along the **vertical direction on the CCD** (but rotating with respect to the solar surface). To do this, only one rotational transformation of the Stokes component is needed (i.e. not rotation of the images themselves). However the angle of rotation changes throughout the day. The angle $\alpha$, measured from the CCD's vertical direction to the demodulated images' positive $Q$, is

$$\alpha = \pm(C + \alpha_{\text{el}} - \alpha_{\text{az}}) \tag{3.1}$$

where the sign depends on the number of reflections between lens and CCD, and $C$ is a constant angle for the optical setup. Determining the sign and the constant $C$ is done by attaching an opaque object across the telescope lens, aligned to vertical, and the re-image the pupil onto the CCD. Two images taken at least a few minutes apart is enough. The direction of rotation of the re-imaged pupil determines the sign, and $C$ is calculated by measuring the angle $\alpha$ in either of the images (or both), and using the telescope coordinates at the time of observation. For convenience it is assumed that $\alpha$ is always positive for counterclockwise rotation of the image's positive $Q$ direction. For imaging through SOUP, the sign is positive and $C \approx -41.7°$ (measured on June 4, 2005).

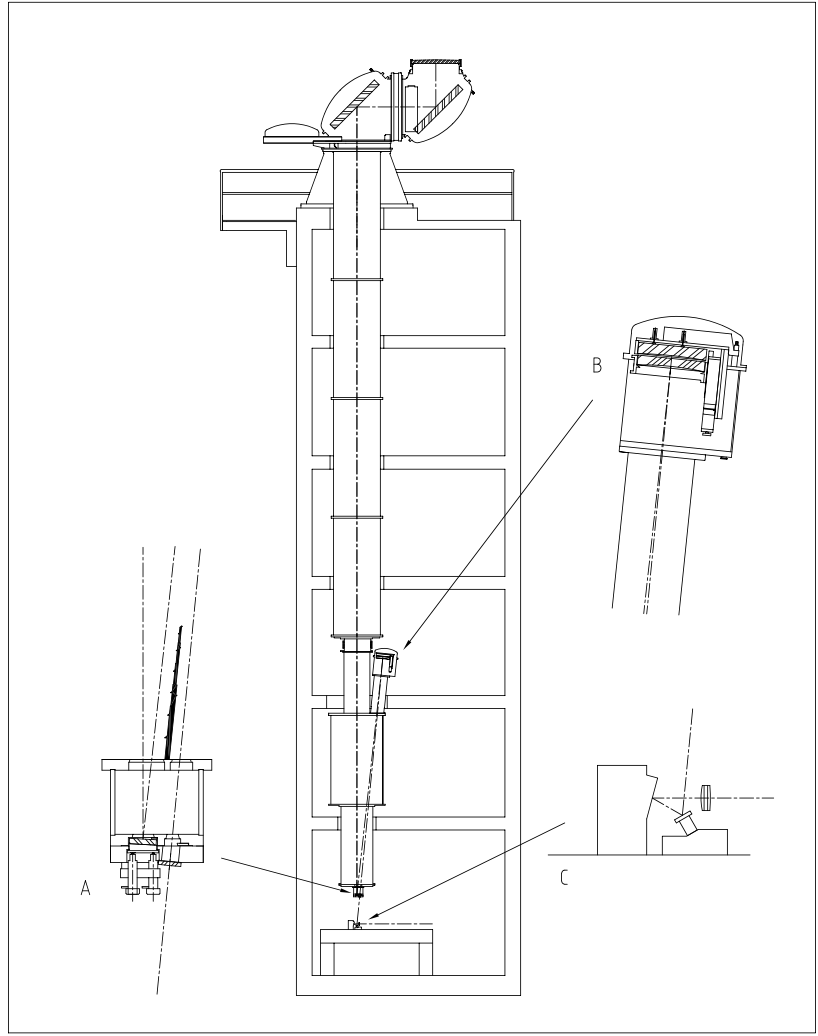

Figure 3.1: Telescope design overview; A=field mirror and field lens, B=Schupmann corrector, C=AO system (tip/tilt-mirror, deformable mirror and re-imaging lens).

Once the angle $\alpha$ can be calculated for any telescope coordinates, rotating positive $Q$ back to vertical is done simply by mutliplying the Stokes vector with the rotational transformation $R(-\alpha)$. Note that it is only $Q$ and $U$ that change. The operation is

$$\boldsymbol{S}_{\text{CCD}} = \mathbf{R}(-\alpha)\boldsymbol{S}_{\text{lens}} \qquad (3.2)$$

If instead a coordinate frame **fixed to the Sun** (positive Q along solar North-South) is desired, the solar tilt angle $\alpha_{\text{t}}$ is needed. This angle is defined as

$$\alpha_{\text{t}} = p - s \qquad (3.3)$$

where $p$ is the angle between the solar meridian and the vertical circle (parallactic angle) measured counterclockwise from solar meridian, and $s$ is the angle between terrestrial North point and solar North on the solar disk, measured counterclockwise from the terrestrial North point. The solar tilt changes over day (it is mainly $p$ that changes), and is conveniently logged by the turret software. Thus at any point, rotating the Stokes vector back to the Sun's coordinate frame is done through the operation[1]

$$\boldsymbol{S}_{\text{Sun}} = \mathbf{R}(-\alpha_{\text{t}})\boldsymbol{S}_{\text{lens}} \qquad (3.4)$$

**Turret model**

The turret includes the main lens and the two turret mirrors.

The coordinate frame for the lens is oriented so that positive $Q$ is along the vertical line, and the matrix is denoted $\mathbf{M}_{\text{main}}$. The two folding mirrors have 45° angle of incidence, and their matrices are $\mathbf{M}_{\text{el}}$ and $\mathbf{M}_{\text{az}}$. The elevation mirror matrix is written in the frame where $Q$ is perpendicular to the plane of incidence. Since this is the same as the $Q$ direction of the lens there is no need for coordinate transformation. However for the azimuth mirror, the vector output from the elevation mirror needs to be transformed to an angle equal to $(\pi/2 - \alpha_{\text{el}})$. It is then necessary to transform the vector output by the azimuth mirror to a frame that is fixed to the Earth's rotational axis (the azimuth mirror rotates, and its frame rotates with it). The angle between the azimuth mirror's positive $Q$ and terrestrial North-South direction is $-\alpha_{\text{az}}$. The whole matrix for the turret can be written as

$$\mathbf{T}_{\text{tur}} = \mathbf{R}(-\alpha_{\text{az}})\mathbf{M}_{\text{az}}\mathbf{R}(\pi/2 - \alpha_{\text{el}})\mathbf{M}_{\text{el}}\mathbf{L}_{\text{main}} \tag{3.5}$$

---

[1]The implementation in software is different. Instead of inverting the telescope matrix, and then doing the rotation, the product of the rotational transformation and the telescope matrix is inverted.

This is the only time/coordinate-dependent part of the telescope matrix.

It should be stressed that the telescope coordinates $\alpha_{\text{el}}$ and $\alpha_{\text{az}}$ are not the same as the telescope target coordinates, because the Schupmann corrector operates a few tenths of a degree off-axis. The rotational difference is large (several degrees) when observing close to the zenith. The easiest way of finding the telescope coordinates is to use the log file that is created by the turret control software. Coordinates are written to it every 30 seconds, and by interpolating these values one can find coordinates for any given time.

**Fixed telescope matrix**

The next part of the telescope, which includes the field mirror, the Schupmann corrector, and the field lens, is independent of pointing. It will be referred to as the fixed telescope matrix, $\mathbf{T}_{\text{fix}}$. First the incoming vector must be transformed to the coordinate frame where positive $Q$ is perpendicular to the plane of incidence of the field mirror. For this purpose we need to know the angle between terrestrial North-South and the field mirror's plane of incidence, denoted $\alpha_{\text{N/S}}$. The field mirror has a 3.125° angle of incidence, and matrix $\mathbf{M}_{\text{f}}$. Next is the Schupmann lens with matrix $\mathbf{L}_{\text{sch}}$. Following the lens is the Schupmann mirror with 0.35° angle of incidence, and matrix $\mathbf{M}_{\text{sch}}$. After that the Schupmann lens is passed once more, this time however the matrix will be different since the light moves in opposite direction. This will be denoted by $\widetilde{\mathbf{L}}_{\text{sch}}$. The last element is the field lens, with matrix $\mathbf{L}_{\text{f}}$. Last, to transform back to the North-South coordinate frame, an additional rotation of $-\alpha_{\text{N/S}}$ is needed. The matrix for the Schupman system and field optics will now be

$$\mathbf{T}_{\text{fix}} = \mathbf{R}(-\alpha_{\text{N/S}})\,\mathbf{L}_{\text{f}}\,\widetilde{\mathbf{L}}_{\text{sch}}\,\mathbf{M}_{\text{sch}}\,\mathbf{L}_{\text{sch}}\,\mathbf{M}_{\text{f}}\,\mathbf{R}(\alpha_{\text{N/S}}) \tag{3.6}$$

It should be pointed out that this matrix is probably very weakly polarizing, due to the small angles of incidence and absence of mechanical stress on the Schupmann lens (stress usually induces birefringence). The exit lens, which is under mechanical stress, is relatively thin so that it should introduce only slight polarization effects.

The complete telescope Müller matrix is

$$\mathbf{T}_{\text{tel}} = \mathbf{T}_{\text{fix}}\,\mathbf{T}_{\text{tur}} \tag{3.7}$$

where the coordinate frame of the incident Stokes vector is fixed to the main lens (positive $Q$ is vertical), and the frame of the outcoming vector has positive $Q$ along terrestrial North-South.

**Simplified telescope model**

As mentioned earlier, the major source of polarization is likely to be the main lens and the two folding mirrors. Therefore a simplified model has been created instead.

Looking at the fixed telescope matrix, one can draw some conclusions about the magnitude of polarization effects. First of all the Schupmann corrector, consisting of one lens and one mirror, has an angle of incidence of only $0.35°$. If values for aluminium ($n = 0.77$ and $k = 6.1$) are used in the *physical* mirror model, the Schupmann mirror is likely to have linear polarization effects of only $10^{-6}$ and retardance effects of $10^{-5}$. Assuming a zero-degree mirror is very reasonable. For the Schupmann lens, which lacks mechanical stress, any polarization effects are believed to be very small. The field mirror, however has an angle of incidence of $3.125°$, and the physical mirror model gives linear polarization of $10^{-4}$ and retardance of $10^{-3}$. The exit lens is believed to have only a slight polarizing effect. Thus the whole fixed telescope matrix could be decently approximated by only a few very weak retarders (field mirror, Schupmann lens and exit lens), and two zero-degree mirrors which will cancel each other. *Since the field mirror is likely the most significant element, we will approximate the fixed telescope matrix with that of one free mirror plus one zero-degree mirror, including the angle of orientation.*

$$\mathbf{T}_\text{fix} \approx \mathbf{M}_0 \ \text{Rot}(\mathbf{M}_\text{f}, \alpha_\text{N/S}) \qquad (3.8)$$

In this simplified fixed telescope matrix, we have only 3 free parameters.

For the turret model, we model the folding mirrors as free metallic mirrors, with identical properties. This gives 2 more parameters.[2] It would of course be possible to assume non-identical folding mirrors, but simulations have showed that the properties of the lens and the elevation mirror are difficult to separate.

The model of the main lens poses a problem. Mechanical stress introduces birefringence in optical elements. The vacuum load is the largest contributing force, expected to give rotationally symmetrical birefringence, increasing with the distance from the center of the lens. However, our measurements clearly show that the load on the lens support is not uniform along the perimeter of the lens and that the birefringence is far from rotationally symmetric. Furthermore, the gravitational stress can cause non-symmetrical birefringence which varies with the telescope elevation angle. It is also possible that there are remaining stress patches from the manufacturing process.

There have been several attempts to model entrance windows of vaccum telescope such as the VTT [6] and the SVST [8]. In both of these cases, the window has been approximated by a linear retarder, and in both cases the estimated retardance has been of the order of a few degrees. However, with a pure vacuum load it is not possible to find a main axis, due to the symmetry of the birefringence. Hence it is unlikely that the effects of the vacuum load can be entirely modelled this way.

---
[2]Note that the angle of incidence is not used in the free metallic mirror model.

Instead a general birefringent window model has been suggested [2]. It is a 5-parameter Müller matrix which can model the total effect of an optical element with arbitrary birefringence (both retardance and axis), and is given in Section B.3.6. This model is used because the images show large fluctuations of the retardance near the edge of the pupil (60° peak, and more than 30° for about 2% of the pupil area [1]).

This matrix has been deduced from a model where locally the window behaves like a linear retarder with arbitrary retardance and axis. The local Müller matrix is

$$\mathbf{M}_{x,y} = \begin{pmatrix} 1 & 0 & 0 & 0 \\ 0 & \cos^2 2\alpha + \sin^2 2\alpha \cos\delta & \sin 2\alpha \cos 2\alpha \left(1 - \cos\delta\right) & -\sin 2\alpha \sin\delta \\ 0 & \sin 2\alpha \cos 2\alpha \left(1 - \cos\delta\right) & \sin^2 2\alpha + \cos^2 2\alpha \cos\delta & \cos 2\alpha \sin\delta \\ 0 & \sin 2\alpha \sin\delta & -\cos 2\alpha \sin\delta & \cos\delta \end{pmatrix} \quad (3.9)$$

The matrix for the whole window, $\mathbf{M}$, is the local matrix averaged over the full aperture[3]. This gives

$$\mathbf{M} = \langle \mathbf{M}_{x,y} \rangle = \begin{pmatrix} 1 & 0 & 0 & 0 \\ 0 & A & B & -C \\ 0 & B & D & E \\ 0 & C & -E & F \end{pmatrix} \quad (3.10)$$

where

$$\begin{aligned} A &= \langle \cos^2 2\alpha + \sin^2 2\alpha \cos\delta \rangle \\ B &= \langle \sin 2\alpha \cos 2\alpha \left(1 - \cos\delta\right) \rangle \\ C &= \langle \sin 2\alpha \sin\delta \rangle \\ D &= \langle \sin^2 2\alpha + \cos^2 2\alpha \cos\delta \rangle \\ E &= \langle \cos 2\alpha \sin\delta \rangle \\ F &= \langle \cos\delta \rangle \end{aligned} \quad (3.11)$$

However, the parameter $F$ is redundant since

$$\begin{aligned} A + D - 1 &= \langle \cos^2 2\alpha \rangle + \langle \sin^2 2\alpha \cos\delta \rangle + \\ &\quad + \langle \sin^2 2\alpha \rangle + \langle \cos^2 2\alpha \cos\delta \rangle - 1 \\ &= 1 + \langle \sin^2 2\alpha \cos\delta + \cos^2 2\alpha \cos\delta \rangle - 1 \\ &= \langle \cos\delta \rangle = F \end{aligned} \quad (3.12)$$

The simplified telescope model would then be

$$\widetilde{\mathbf{T}}_{\text{tel}} = \mathbf{M}_0 \operatorname{Rot}(\mathbf{M}_{\text{f}}, \alpha_{\text{f}}) \mathbf{R}(-\alpha_{\text{az}}) \mathbf{M}_{\text{az}} \mathbf{R}(\pi/2 - \alpha_{\text{el}}) \mathbf{M}_{\text{el}} \mathbf{L}_{\text{main}} \quad (3.13)$$

---

[3]Note that we assume on-axis observations. Also, the PSF of the lens is distorted, making this a most undesirable effect.

with $\mathbf{M}_0$ being a zero-degree mirror, $\mathbf{M}_\mathrm{f}$ a free mirror (rotated by angle $\alpha_\mathrm{f}$), $\mathbf{R}(-\alpha_\mathrm{az})$ and $\mathbf{R}(\pi/2 - \alpha_\mathrm{el})$ are rotational transformations, $\mathbf{M}_\mathrm{az}$ and $\mathbf{M}_\mathrm{el}$ are identical free mirrors, and finally $\mathbf{L}_\mathrm{main}$ being the general birefringent window model.

The free mirror model used here is more attractive than the physical model, because it uses only 2 parameters, instead of 2 parameters and angle of incidence. Furthermore it is "free" in the sense that the physical constraints are not present. Thus a mirror that deviates slightly from the physical model, e.g. due to contaminations or similar, is better modelled by a free mirror. This choice is the same as for the SVST [8], but different from e.g. the VTT[6].

## 3.2 Calibration of the SST polarization model

Measurements have been made at several occasions, but only the ones from May 8, 2005, have good polarimeter calibrations. They are therefore used in this document.

The measurement setup uses the 1-meter polarizer in front of the telescope main lens. It is rotated in steps of 22.5° by computer control, and at each point a polarimetric observation is made. The elapsed time between consecutive measurements is not critical, however the time required for *one* measurement is, because the modulation is assumed to be for a constant Stokes vector, something which is not true since the telescope rotates the Stokes vector. The time needed for one polarimetric observation was in this case around 8 seconds.

Each observation is defined by

1. 20 modulated images, 5 for each state of modulation

2. the time of each measurement, as written in the image file headers (the average of all 20 images is used)

3. the angle of the 1-meter polarizer (measured from the vertical line, counterclockwise as seen from the observer receiving the light)

To extract the telescope coordinates (elevation and azimuth) from the time, the log file created by the turret software is used (it is *not* sufficient to assume that the telescope coordinates are identical to the Sun's coordinates due to the off-axis design).

The optical setup for the observations was

1. 40″ pinhole (4 mm) at telescope exit window, to smooth out solar structures.

2. Reflective ND filter 3.0, slightly tilted, and absorbing ND filter 0.5.

3. Calibrational linear polarizer #2 in 0°/90° mount, with an angle error of 0.9° relative to North-South measured counterclockwise as seen from the observer receiving the light, when the mounts in in 0° position. This was determined by adjusting the telescope coordinates so that the image of the 1-meter polarizer vertical beam coincided with the indicated axis of the calibrational polarizer, which happened at elevation 0° and azimuth 90.9°.

4. Quarter-wave plate on Newport rotator stage. The quarter-wave plate had been manually aligned with the calibrational polarizer to within about 10°.

5. AO system.

6. Sliding mirror, reflecting light back to a track next to the AO system.

7. LCVRs, including heat shield. LCVR 1 at 0° and LCVR 2 at 45°.

8. Linear polarizer #1 (analyzer), flipped, at approximately 0° (vertical).

9. Re-imaging lens with $f = 260$ mm.

10. Spectrograph filter, 6302 Å.

11. CCD camera XIII, at 0 dB gain.

In these observations the pupil was re-imaged on the CCD by the $f = 150$ mm lens. This method has several advantages:

1. When the properties of the telescope polarization model have been determined, the properties of the lens can be fully determined point by point across the pupil [1], in terms of birefringence (retardance and local optical axis). This can also serve as a check of the assumed model of the lens.

2. The accuracy of the telescope azimuth and elevation angles can be verified as regards to their combined effects on pupil rotation at the focal plane

3. The accuracy of the 1-meter polarizer wheel angles can be verified

The errors in telescope coordinates and 1-meter polarizer angle are treated in Section 3.2.1.

Due to NLR, each polarimetric observation is repeated twice at different exposure times—first 100 ms then 50 ms—so that it is possible to use the double exposure method for correction. For the "Frames" setting in the camera software, a value of 5 was used. However, due to a bug in the camera software, the first frame acquired after changing exposure actually uses the previous exposure. By having 4 reliable frames, one can average out the odd/even effect of the camera shutter, and also get an idea of the SNR.

| Image # | Exposure (ms) | Mode | Polarizer angle |
|---|---|---|---|
| 201–205 | 50, 100, 100, 100, 100 | 1 | 0° |
| 206–210 | 100, 100, 100, 100, 100 | 2 | 0° |
| 211–215 | 100, 100, 100, 100, 100 | 3 | 0° |
| 216–220 | 100, 100, 100, 100, 100 | 4 | 0° |
| 221–225 | 100, 50, 50, 50, 50 | 1 | 0° |
| 226–230 | 50, 50, 50, 50, 50 | 2 | 0° |
| 231–235 | 50, 50, 50, 50, 50 | 3 | 0° |
| 236–240 | 50, 50, 50, 50, 50 | 4 | 0° |
| 241–245 | 50, 100, 100, 100, 100 | 1 | 22.5° |
| 246–250 | 100, 100, 100, 100, 100 | 2 | 22.5° |
| ... | ... | ... | ... |
| 876-880 | 50, 50, 50, 50, 50 | 4 | 360° |
| 881-885 | 50, 100, 100, 100, 100 | 1 | 0° |
| ... | ... | ... | ... |

Table 3.1: Image aquisition pattern for May 8 observations.

To clarify, the frames acquired follows the pattern shown in table 3.1. The full description is in the observation log (`obs_notes.txt`) which can be found together with the data.

### 3.2.1 Verification of telescope coordinates and 1-meter polarizer rotation

The azimuth and elevation coordinates of the telescope determine the relation between the telescope's input and output Stokes vector. During the telescope polarization measurements, the angle of the 1-meter polarizer determines the input Stokes vector. To ensure high accuracy, both telescope coordinates and the 1-meter polarizer angle must be accurate down to a few tenths of a degree.

The rotation angle of the 1-meter polarizer as seen by the CCD camera is given by

$$\alpha = \pm(C + \alpha_{\text{el}} - \alpha_{\text{az}} - \theta) \qquad (3.14)$$

where the sign depends on the re-imaging optics used, $C$ is a constant, $\alpha_{\text{el}}$ and $\alpha_{\text{az}}$ is the elevation and azimuth coordinates of the telescope, and $\theta$ is the rotation angle of the 1-meter polarizer relative to positive $Q$[4]

---

[4]Finding $C$ can be done by attaching a straight opaque object vertically in front of the lens, using a lead-line as reference. By re-imaging the pupil on the CCD and recording the telescope's coordinates at two occasions, $C$ and the sign of rotation can be calculated.

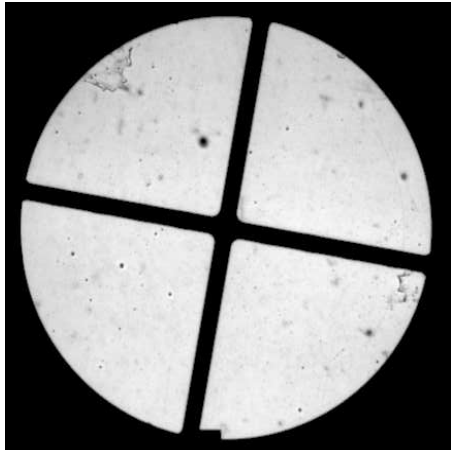

Figure 3.2: Demodulated $I$ image of the telescope pupil, showing the cross-like support of the 1-meter polarizer.

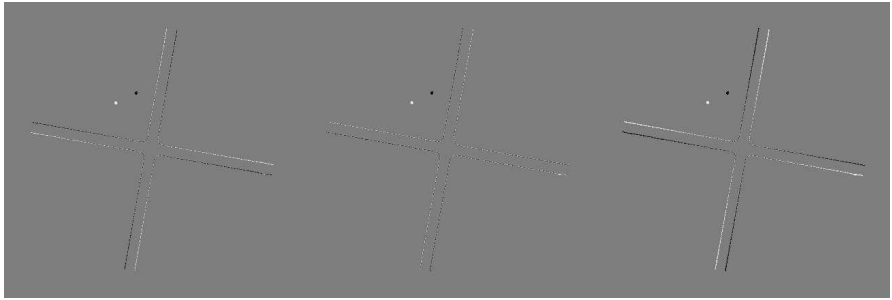

Figure 3.3: Cross-correlation between reference image and another image, rotated to $-0.1°$, $0.2°$, and $0.4°$.

The image of the 1-meter polarizer's cross-like support, shown in figure 3.2, allows us to verify the telescope coordinates and the 1-meter polarizer angle. The first image was used as reference, and each image was converted to a binary mask equal to unity within the cross and zero for other pixels. Each image was then rotated back with the expected angle $\alpha$, and aligned with the reference image. Next, each image was rotated by $-0.2°$, $0.1°$ and $0.4°$ from its nominal position (figure 3.3), and at each position the sum of squares of the difference between the rotated image and reference image was calculated, forming a 3-point error function. The error function was interpolated with a $2^{nd}$ order polynomial in order to find the estimated error in the angle $\alpha$ for each image.

The result is shown in figure 3.4, together with average angle error for each full rotation of the 1-m polarizer. This demonstrates trends in the tele-scope coordinate errors of $\pm 0.1°$ over the whole day. There was no evidence of any systematic angle error for the 1-meter polarizer. Trends were clearly seen during each rotation, but differed from one rotation to another. With these results, it can be assumed that the errors should have very little impact on the telescope polarization model.

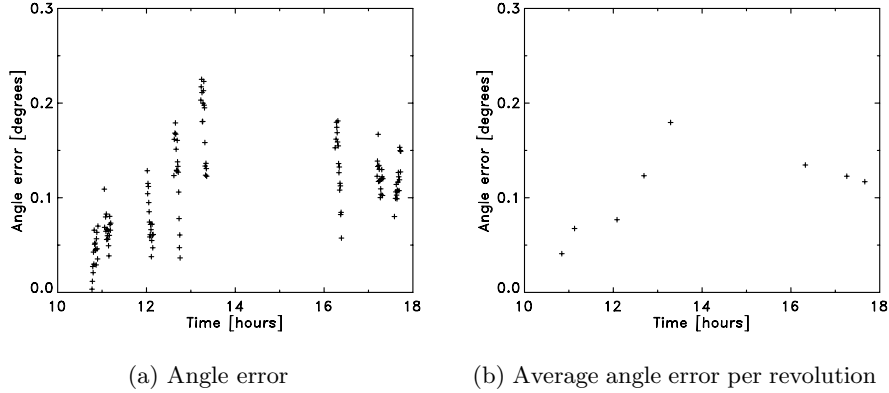

| (a) Angle error | (b) Average angle error per revolution |

Figure 3.4: Measured angle error for 1-meter polarizer cross in images from May 8, 2005.

### 3.2.2 Modulation matrix calibrations

Calibrations were made at 09:41, 15:27 and 18:47 UT, so that it is possible to see whether the modulation is stable over time. At each occasion, two matrix calibrations of different exposures (100 ms and 50 ms) were made. Polarimetric observations of the calibrational polarizer were made in conjunction with the matrix calibrations, and also these were made at both 100 ms and 50 ms exposure. These are used for finding the correct value of the retardance of the calibrational quarter-wave plate. Note that none of the methods for non-linearity compensation was used in these calibrations.

The first 100 ms calibration gives the modulation matrix

$$\mathbf{P} = \begin{pmatrix} 0.9922 & 0.6316 & 0.3657 & -0.6943 \\ 0.9882 & -0.1444 & 0.7688 & 0.6168 \\ 0.9923 & -0.8490 & -0.3007 & -0.4261 \\ 1.0000 & 0.4626 & -0.8088 & 0.4221 \end{pmatrix} \quad (3.15)$$

with modulation efficiencies

$$\begin{pmatrix} \epsilon_\mathrm{I} & \epsilon_\mathrm{Q} & \epsilon_\mathrm{U} & \epsilon_\mathrm{V} \end{pmatrix} = \begin{pmatrix} 0.991 & 0.581 & 0.606 & 0.552 \end{pmatrix} \quad (3.16)$$

The parameters of the quarter-wave plate were determined to

$$\delta_{\mathrm{QW}} = 1.5763 \text{ radians } (90.32°) \quad (3.17)$$
$$D_\alpha = 0.1575 \text{ radians } (9.022°) \quad (3.18)$$

using observations of the linear polarizer at approximately $0°$, $-22.5°$, $-45°$, $-67.5°$ and $-90°$, weighted by 0, 1, 2, 3, and 4 in the $\chi^2$ function (the error in retardance does not contribute to an error in degree of polarization for the first measurement, hence the weight 0). The intensity increase was assumed to be 0, even though it is likely that it was actually a few percent.

The calibration at **15:27** gave the modulation matrix

$$\mathbf{P} = \begin{pmatrix} 0.9933 & 0.6290 & 0.3678 & -0.6905 \\ 0.9887 & -0.1642 & 0.7666 & 0.6143 \\ 0.9966 & -0.8475 & -0.3109 & -0.4365 \\ 1.0000 & 0.4605 & -0.8015 & 0.4283 \end{pmatrix} \quad (3.19)$$

and quarter-wave plate parameters

$$\delta_{\mathrm{QW}} = 1.5739 \text{ radians } (90.18°) \quad (3.20)$$
$$D_\alpha = 0.1625 \text{ radians } (9.310°) \quad (3.21)$$

The calibration at **18:47** gave the following modulation matrix

$$\mathbf{P} = \begin{pmatrix} 0.9744 & 0.6252 & 0.3937 & -0.6622 \\ 1.0000 & -0.1899 & 0.7788 & 0.6255 \\ 0.9867 & -0.8381 & -0.3100 & -0.4206 \\ 0.9945 & 0.4442 & -0.7813 & 0.4501 \end{pmatrix} \quad (3.22)$$

and quarter-wave plate parameters

$$\delta_{\mathrm{QW}} = 1.5655 \text{ radians } (89.70°) \quad (3.23)$$
$$D_\alpha = 0.1640 \text{ radians } (9.40°) \quad (3.24)$$

Since there are some changes in the quarter-wave parameters over time, it is not sure that the change in modulation is real or just due to errors in the quarter-wave plate parameters. However, the measured modulation of V is very insensitive to those errors, suggesting that the change in modulation is in fact real. If the same values of retardance and angle offset are used, the modulation is still different. However the first and last calibration are not reliable due to rapid change in intensity. This was also indicated by the $\chi^2$ values. Therefore only the calibration at 15:27 is believed to be reliable.

### 3.2.3 Telescope polarization data

For the telescope polarization data, the calibration from 15:27 is used for demodulation.

| Revolution | Started (UT) | Ended (UT) |
|---|---|---|
| 1 | 10:46 | 10:54 |
| 2 | 11:03 | 11:12 |
| 3 | 12:01 | 12:09 |
| 4 | 12:37 | 12:45 |
| 5 | 13:13 | 13:21 |
| 6 | 16:15 | 16:23 |
| 7 | 17:12 | 17:19 |
| 8 | 17:35 | 17:44 |

Table 3.2: Starting time of each revolution of 1-meter polarizer.

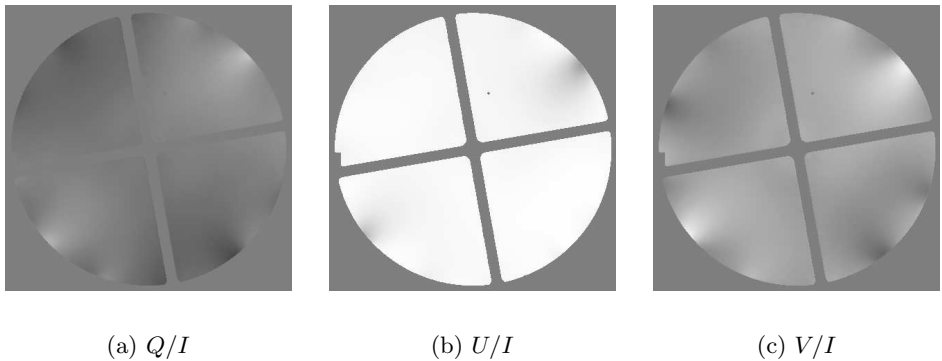

(a) $Q/I$    (b) $U/I$    (c) $V/I$

Figure 3.5: Polarization images of telescope pupil, with 1-meter polarizer wheel (transmitting axis is almost horizontal). Images are scaled from $-1$ to $1$. The stress-induced birefringence along the edges is very clear. Note that the polarization occuring after the lens introduces cross-talk between images (but not *within* images).

In total eight full revolutions of the large polarizer were made. Note that each revolution starts with $0°$ and ends with $360°$, i.e. the last measurement is redundant[5]. The time of each revolution is given in table 3.2. The Stokes components for the first revolution is shown in figure 3.6(a). It has also proved useful to check the degree of polarization for the measured data, because a value larger than 1 indicates demodulation errors (like in the modulation matrix calibration). The degree of polarization is shown in figure 3.6(b).

The interesting feature in the data is the $V/I$ component. It is purely created by cross-talk from the other components, since the 1-meter polarizer by itself produces very little $V/I$[6]. Note that the $Q/I$ component starts at

---

[5] However it cannot directly be used as consistency check with the first measurement since the telescope moves too much between the two measurements.

[6] Measurements in lab indicate less than 0.5% $V/I$ is produced.

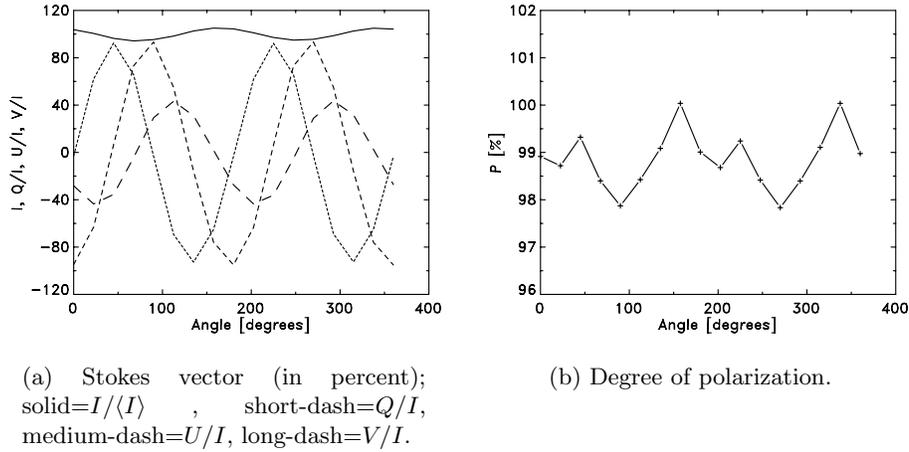

(a) Stokes vector (in percent); solid=$I/\langle I\rangle$, short-dash=$Q/I$, medium-dash=$U/I$, long-dash=$V/I$.

(b) Degree of polarization.

Figure 3.6: Polarization data for first revolution of the 1-meter polarizer.

0 only by chance; the phase of the components varies with the telescope coordinates. The degree of polarization varies periodically with 180°. It is possible that this is a true variation, or that it is just errors in the demodulation matrix. However it is never significantly above 1, so the data is not unreasonable. All other revolutions look in principle the same, with a slight amplitude and/or phase shift throughout the day, and they are therefore not shown.

Looking at the polarization images in figure 3.5, it is apparent that the lens has a great deal of stress-induced birefringence, from the vacuum load.

### 3.2.4 Model fitting

The parameters of the simplified telescope model were fitted to the telescope polarization data 100 times, and each time with a randomly (within reasonable limits) selected starting solution. Table 3.3 shows the statistical results, which were quite interesting; *the fitted solution is in principle exactly the same each time.* This was not the case with SVST, as described in [8], where the solutions were more spread out. Note that some of the parameters have multiple equivalent values. In those cases they have been reduced to only one equivalent solution.

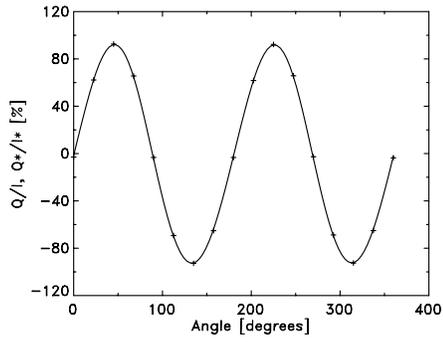

(a) $Q/I$

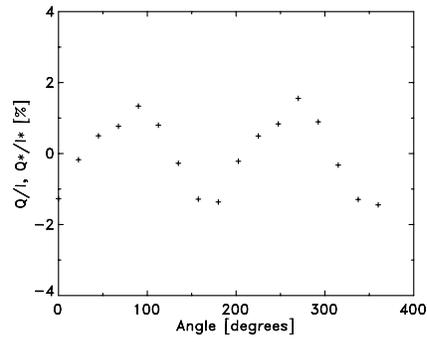

(b) $Q/I$ residual

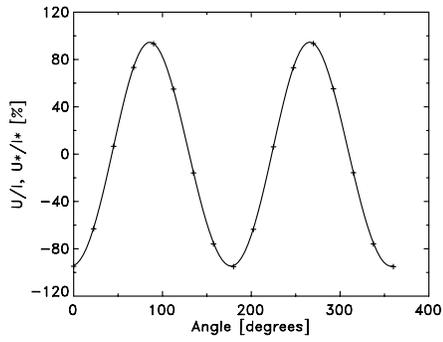

(c) $U/I$

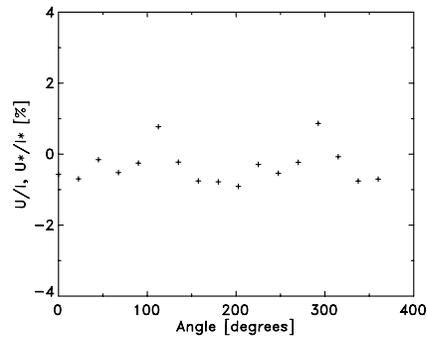

(d) $U/I$ residual

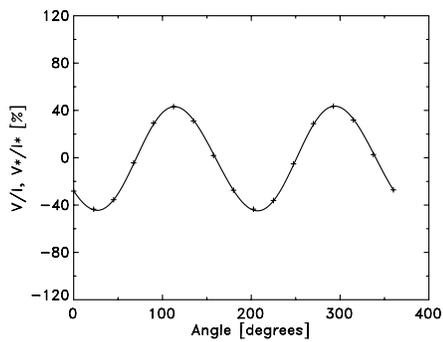

(e) $V/I$

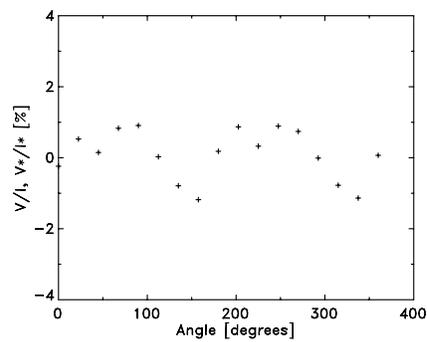

(f) $V/I$ residual

Figure 3.7: Measured data (crosses) and fitted data (solid) for first revolution of the 1-meter polarizer. The residual errors are also plotted separately for visibility.

| Parameter | Symbol | Mean | Variation |
|---|---|---:|---|
| **Main lens** | | | |
| A | $A$ | $9.809484 \cdot 10^{-1}$ | $\pm 5 \cdot 10^{-7}$ |
| B | $B$ | $8.4169 \cdot 10^{-4}$ | $\pm 8 \cdot 10^{-7}$ |
| C | $C$ | $-2.64989 \cdot 10^{-2}$ | $\pm 8 \cdot 10^{-6}$ |
| D | $D$ | $9.890081 \cdot 10^{-1}$ | $\pm 3 \cdot 10^{-6}$ |
| E | $E$ | $-4.7782 \cdot 10^{-2}$ | $\pm 6 \cdot 10^{-5}$ |
| **Folding mirrors** | | | |
| polarization | $R_\text{m}$ | $9.29439 \cdot 10^{-1}$ | $\pm 2 \cdot 10^{-5}$ |
| retardance | $\delta_\text{m}$ | $16.8994°$ | $\pm 2 \cdot 10^{-3}$ |
| **Field mirror** | | | |
| polarization | $R_\text{f}$ | $9.99995 \cdot 10^{-1}$† | $\pm 5 \cdot 10^{-5}$ |
| retardance | $\delta_\text{f}$ | $0.58784°$† | $\pm 4 \cdot 10^{-4}$ |
| axis | $\alpha_\text{f}$ | $32.94°$† | $\pm 2 \cdot 10^{-1}$ |
| $\chi^2$ | n/a | $2.987595 \cdot 10^{-5}$ | $\pm 8 \cdot 10^{-11}$ |

Table 3.3: Fitted model parameters. Note how extremely well-defined all parameters are. † indicates that several equivalent solutions were reduced to one.

The best fit was

$$A = 9.809483 \cdot 10^{-1} \tag{3.25}$$
$$B = 8.415676 \cdot 10^{-4} \tag{3.26}$$
$$C = -2.650319 \cdot 10^{-2} \tag{3.27}$$
$$D = 9.890090 \cdot 10^{-1} \tag{3.28}$$
$$E = -4.776245 \cdot 10^{-2} \tag{3.29}$$
$$R_\text{m} = 9.294432 \cdot 10^{-1} \tag{3.30}$$
$$\delta_\text{m} = 16.89882° \tag{3.31}$$
$$R_\text{f} = 1.000008 \tag{3.32}$$
$$\delta_\text{f} = 0.5880293° \tag{3.33}$$
$$\alpha_\text{f} = 32.97094° \tag{3.34}$$
$$\chi^2 = 2.987587 \cdot 10^{-5} \tag{3.35}$$

using the same symbols as in table 3.3. The reproduced data is shown in figure 3.7. Hence, the Müller matrix of the lens would be

$$\mathbf{L}_\text{main} = \begin{pmatrix} 1 & 0 & 0 & 0 \\ 0 & 0.9809 & 0.0008 & 0.0265 \\ 0 & 0.0008 & 0.9890 & -0.0478 \\ 0 & -0.0265 & 0.0478 & 0.9700 \end{pmatrix} \tag{3.36}$$

The local properties of the lens have also been determined[1], using the pupil images demodulated in the coordinate frame at the exit side of the lens (after all parameters of the telescope polarization model have been determined). Using one revolution of the 1-meter polarier for first determining the local retardance and calibration, *then* calculating the lens parameters gives, preliminary,

$$A = 0.9848 \quad (3.37)$$
$$B = 0.0047 \quad (3.38)$$
$$C = -0.0279 \quad (3.39)$$
$$D = 0.9935 \quad (3.40)$$
$$E = -0.0463 \quad (3.41)$$

With these parameters, the Müller matrix of the lens is

$$\widetilde{\mathbf{L}}_{\text{main}} = \begin{pmatrix} 1 & 0 & 0 & 0 \\ 0 & 0.9848 & 0.0047 & 0.0279 \\ 0 & 0.0047 & 0.9935 & -0.0463 \\ 0 & -0.0279 & 0.0463 & 0.9783 \end{pmatrix} \quad (3.42)$$

All five parameters are fairly consistent with those given in this report.

The Müller matrix of the telescope during May 9, 2005, is shown in figure 3.8.

### 3.2.5 Telescope model validation for direct sunlight

To test the telescope model, polarimetric measurements for direct sunlight were made on May 18 and 23, 2005. The data from May 23 is used here.

The spatially averaged light of the solar surface can be assumed unpolarized. Thus any polarization measured comes from the telescope itself. This also more accurately represents the real conditions under which the telescope model is used, i.e. in regular observations *without* the 1-meter polarizer attached to the telescope.

The calibration at 15:17 gave the modulation

$$\mathbf{P} = \begin{pmatrix} 0.9891 & 0.5981 & 0.3798 & -0.6932 \\ 0.9863 & -0.1229 & 0.7688 & 0.6157 \\ 0.9937 & -0.8555 & -0.3026 & -0.4171 \\ 1.0000 & 0.4668 & -0.7877 & 0.4005 \end{pmatrix} \quad (3.43)$$

and quarter-wave plate parameters

$$\delta_{\text{QW}} = 1.5738 \text{ radians } (90.17°) \quad (3.44)$$
$$D_\alpha = 0.1599 \text{ radians } (9.16°) \quad (3.45)$$

The measured data versus predicted data is shown in figure 3.9. The last measurements were made very late in the day when the Sun was only a few degrees above the horizion.

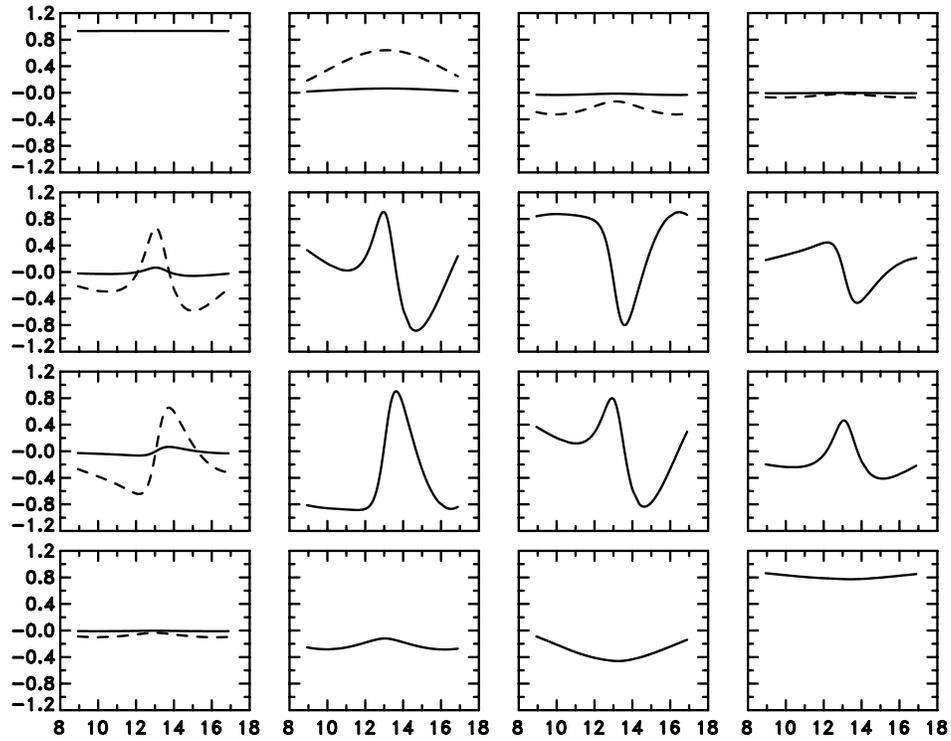

Figure 3.8: Calculated telescope Müller matrix during May 9, 2005, for 630.2 nm. The solid line is the element value, and the dashed line is the element multiplied by 10 (for visibility). The position of each plot corresponds to the position of the element in the matrix, i.e. the first row of plots is the first row of elements in the Müller matrix, and equivalently for the columns.

Looking at those plots, it is evident that the error in predicted $I$ to $Q$ cross-talk is overestimated by $+4\%$. The $I$ to $U$ cross-talk is also overestimated by $+4\%$, but cannot be fully modelled this way. However adding a cross-talk from $Q/I$ of 2% reduces the error. The $V/I$ error is mainly in the form of an offset of 0.16%. Figure 3.10 shows the measured and replicated data when these adjustments are applied.

The source of the errors is not currently known. It is essentially unchanged if the data is demodulated using the matrix from May 8, 2005. However, even with this error the accuracy of the model is excellent.

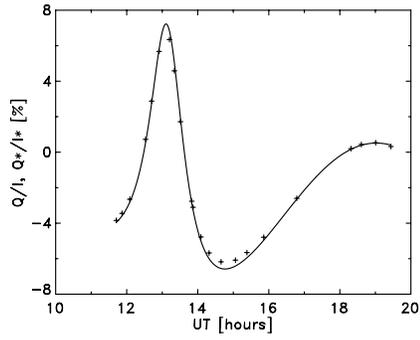
(a) $Q/I$

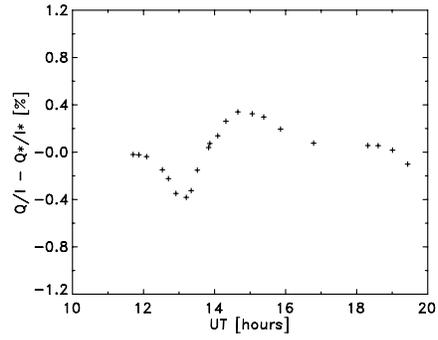
(b) $Q/I$ error

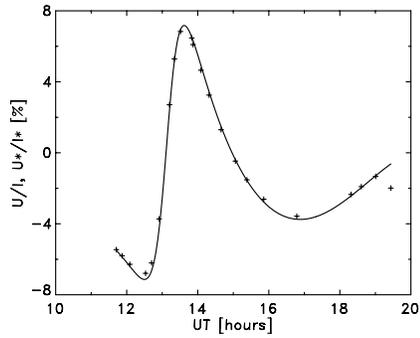
(c) $U/I$

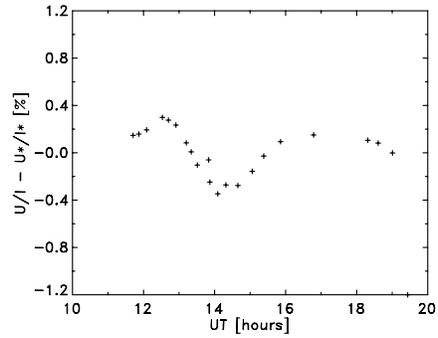
(d) $U/I$ error

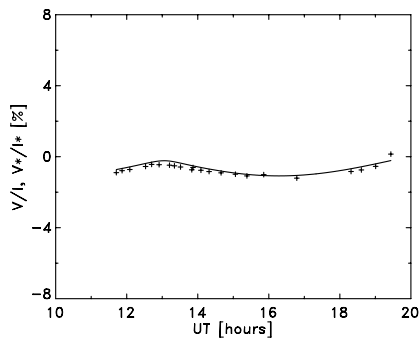
(e) $V/I$

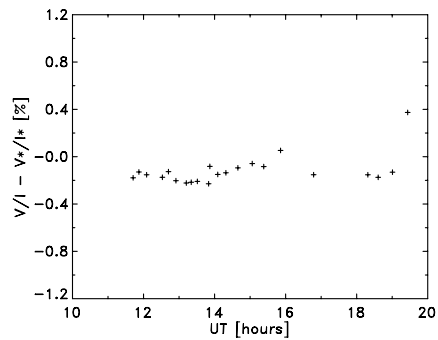
(f) $V/I$ error

Figure 3.9: Measured versus predicted telescope polarization for unpolarized light. The last few points have larger errors most likely because the Sun was only a few degrees above the horizon for these measurements.

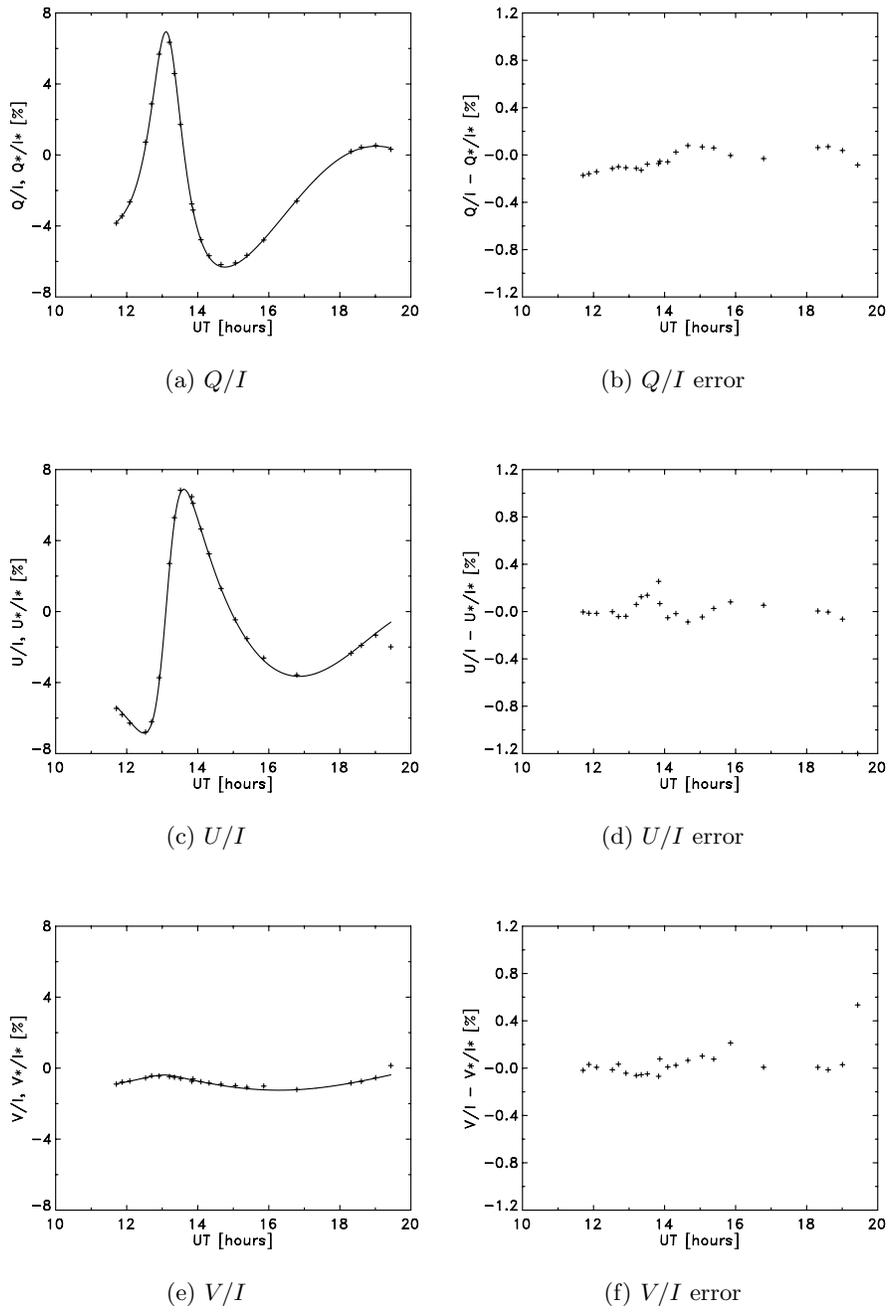

Figure 3.10: Measured versus predicted telescope polarization with adjustments of *predicted* data. $Q/I$ has an amplitude change of $-4\%$. $U/I$ has amplitude change of $-4\%$, and 2% cross-talk from $Q/I$ added. $V/I$ has an offset of $-0.16\%$ added.

# Chapter 4

# First light with a Stokes imaging polarimeter

Some of the first fully demodulated and telescope compensated polarimetric images are shown here.[1] They were acquired on June 3, 2005, and show the irregular active region AR0772, imaged through SOUP tuned to wavelength 630.2 nm (Fe I) at $-80$ mÅ from line center (blue wing). Images were acquired at 11:03:13 UT, and the region had Stonyhurst coordinates S $17.0°$ and E $18.15°$. For each modulation, three images were acquired, flat-fielded, and restored using MFBD[2].

Figure 4.1 shows the demodulated $I$ component. Figure 4.2, 4.3, and 4.4 show the $Q/I$, $U/I$ and $V/I$ components in two versions: one without telescope compensation ("raw"), and one with telescope compensation ("compensated"). Note that an estimated residual $I$ cross-talk has been subtracted from all images (after demodulation and telescope compensation), using the assumption of zero average polarization in quiet regions on the surface. The operation on each Stokes componens $S_k$ is then

$$\tilde{S}_k = S_k - \frac{\langle S_{k,q} \rangle}{\langle S_1 \rangle} \cdot S_1, \quad k = 2, 3, 4 \tag{4.1}$$

where $S_{k,q}$ is a quiet region in the $S_k$ component. The estimated cross-talk from $I$ was below 0.5% for all components.

The calculated telescope matrix at the time of observation was

$$\mathbf{M}_{\text{tel}} = \begin{pmatrix} 0.931 & 0.051 & -0.028 & -0.005 \\ -0.045 & -0.326 & 0.824 & 0.243 \\ -0.038 & -0.817 & -0.224 & -0.338 \\ -0.009 & -0.246 & -0.345 & 0.801 \end{pmatrix} \tag{4.2}$$

---

[1]Courtesy of O. Khomenko and M. Collados, Instituto de Astrofísica de Canarias, Spain.

[2]Multi-frame blind deconvolution.

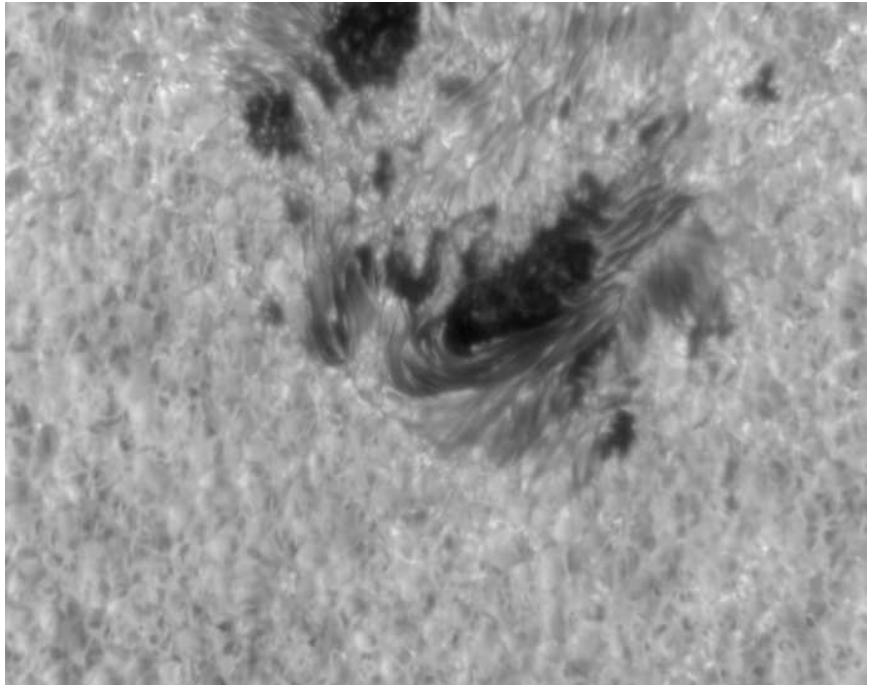

Figure 4.1: Image of active region AR0772, taken at SST on June 3, 2005. Instrument used is SOUP at 630.2 nm, 80 mÅ into blue wing.

The significant cross-talk here is 24% from $V$ into $Q$, $-34$% from $V$ to $U$, and a few % crosstalk from $I$ into $Q$ and $U$. This is easily seen in the images, when comparing with the $V/I$ image. $Q/I$ has a positive $V/I$ cross-talk, and hence the umbra is slightly brighter than the surrounding quiet regions. Hints of faculae can be seen. $U/I$ has essentially the same cross-talk but with a change of sign, making the overlayed $V/I$ image negative. There is also a significant cross-talk from $Q$ and $U$ into $V$ of $-24$% and $-35$% respectively, which mainly affects the appearance of the penumbra in the $V/I$ image.

The appearance of the compensated images is reasonable. Basically the $V/I$ component is strongest in the umbra (center), because the magnetic field is approximately along the LOS (line of sight) for a sunspot this close to disk center. In the penumbra, the direction of the field is gradually moving away from the LOS, creating less $V/I$ and more $Q/I$ and $U/I$. Regions with strong (positive or negative) $Q/I$ has little $U/I$, and vice versa. Also, there are no hints of faculae in the compensated $Q/I$ and $U/I$ images, as opposed to the uncompensated images which show clear traces of faculae (due to the cross-talk from $V$).

The angle of positive $Q$ on the CCD was determined using the pupil re-imaging method. For the expression of $\alpha$ in equation 3.1, the sign was determined to be positive, since $\alpha$ decreased as $(\alpha_{\mathrm{el}} - \alpha_{\mathrm{az}})$ decreased. $C$ was then estimated to $-41.7°$. This gives an angle of $77°$ clockwise for positive $Q$ in these images. This agrees with some of the structures in the penumbra, i.e. the filaments along $77°$ in the $I$ image have also a strong $Q/I$ signal.

The seeing-induced cross-talk is much too high in mainly the $Q/I$ and $U/I$ images, even though MFBD image restoration and destretching has been used. However, the telescope model seems accurate for this particular observation. Completely confirming the model would need spectropolarimetric observations of symmetrics spots close to disk center and limb. The magnetic field of this spot is much too irregular to look for inconsistencies in the telescope compensation without further data analysis, which is beyond the scope of this work.

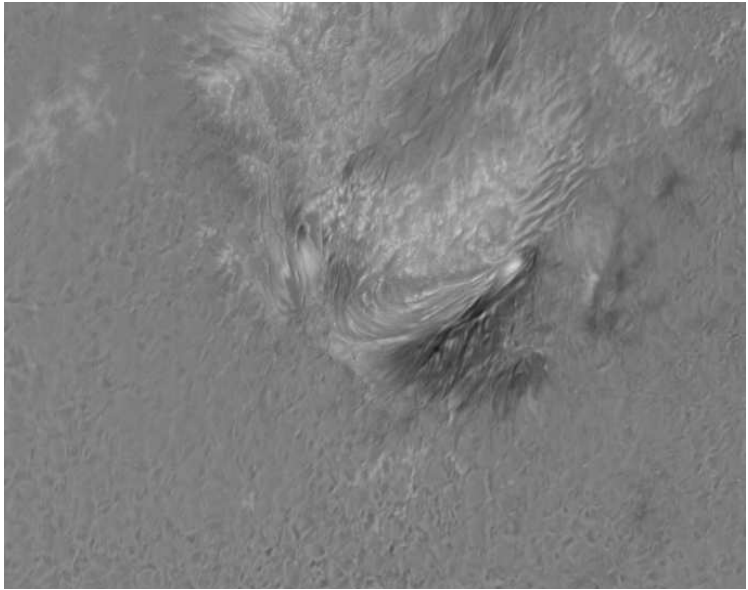

(a) $Q/I$, raw

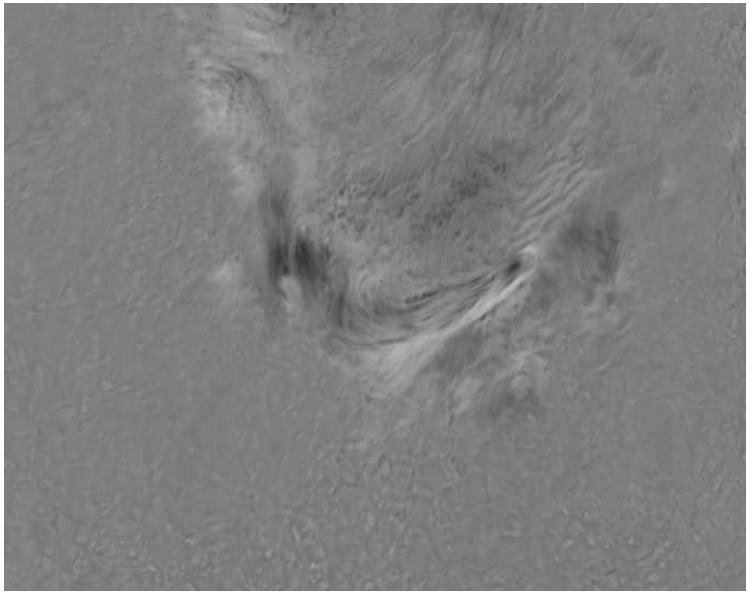

(b) $Q/I$, telescope compensated

Figure 4.2: $Q/I$ component for AR0772, June 3, 2005. Scaling is from $-0.25$ to $+0.25$.

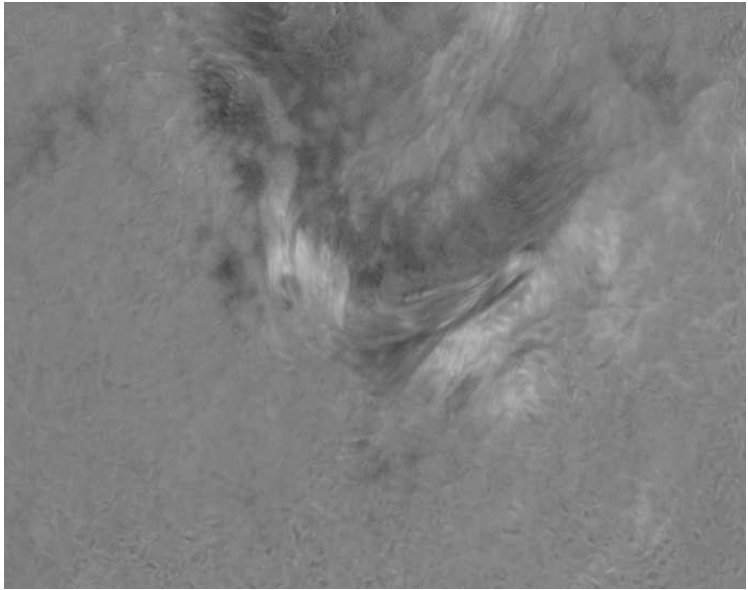

(a) $U/I$, raw

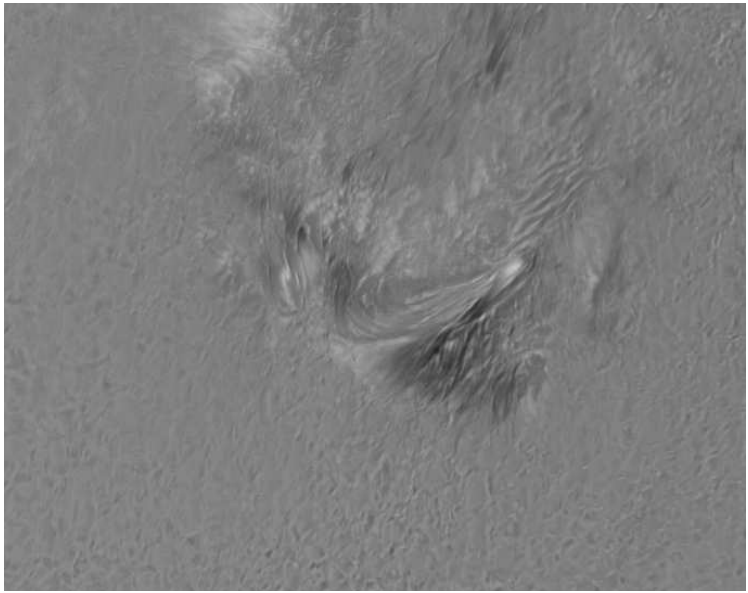

(b) $U/I$, telescope compensated

Figure 4.3: $U/I$ component for AR0772, June 3, 2005. Scaling is from $-0.25$ to $+0.25$.

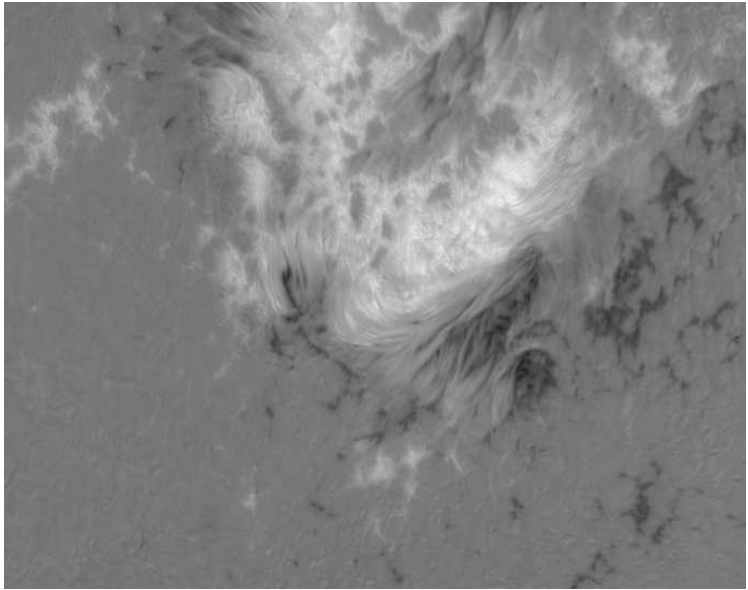

(a) $V/I$, raw

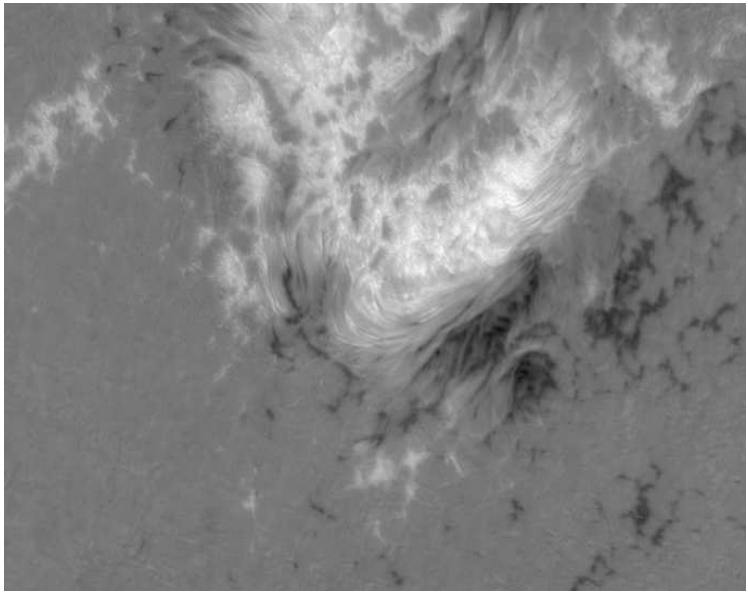

(b) $V/I$, telescope compensated

Figure 4.4: $V/I$ component for AR0772, June 3, 2005. Scaling is from $-0.25$ to $+0.25$.

## Chapter 5

# Conclusions

The SST polarization properties have been succesfully modelled. We have demonstrated that the calibration data obtained with a rotating 1-meter polarizer in front of the lens leads to a polarization model with uniquely defined parameters and with small residual errors in the measured calibration data as compared to model predictions. The model has been independently verified by comparing observed and predicted $I$ to $Q$, $U$ and $V$ cross-talk, with excellent results.

The success of this model can be attributed to several factors:

- Accurate polarimeter calibration, where in particular the severe effects of non-linearities in the CCD response have been investigated and reduced.

- Accurate definition of the positive $Q$-axis and mapping of this axis from the telescope to the observing room.

- The generalized birefringence model of the 1-meter lens, allowing for arbritary variations of the retardance and orientation angle across the lens. This is particularily important with large vacuum windows, for which large retardance near the supporting O-ring is always expected and for which inadequate such support will create locally enhanced retardance that requires characterization by a fairly complex lens Müller matrix.

- Well known azimuth and elevation angles as well as accurately controlled rotation of the 1-meter polarizer.

The main source of error for polarization measurements with the SST presently appears to be inadequate stability of the polarimeter modulation matrix. It is presently not known whether this is due to inadequate temperature stabilization or simply noise in the voltages applied to the LCVRs, but this problem clearly needs to be resolved. A related issue is whether the retardance of the calibration retarder can actually change by a few tenths of a degree, as some measurements possibly suggest, due to daily temperature variations in the observing room.

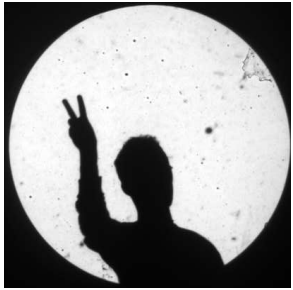

Figure 5.1: The author standing in front of the 1-meter lens of the SST, which is re-imaged on a CCD camera using the same setup as for the telescope polarization measurements.

A number of refinements of the model are possible, but it is not obvious that these need to be implemented: The turret pointing model contains information about alignment errors of the azimuth and elevation axes as well as alignment errors of the mirrors. Feeding this information into the polarization model should allow improved accuracy of the model. The light passing through the entire optical system is not collimated but is actually a converging beam with angles of incidence varying more than $\pm 1°$ from the $f/21$ 1-meter lens and from subsequent re-imaging optics.

Presently, modulation matrix calibration requires nearly twenty minutes of observing time to complete. With more stable LCVRs, such calibrations will not need to be made on a daily basis, but it may still be worth considering rotation of the calibration retarder in increments of ten, instead of five, degrees to speed up this calibration. According to simulations, this will in principle not reduce the accuracy in the modulation matrix.

Re-imaging the 1-meter lens on the CCD used for calibrations has lead to the recognition of the importance of variations in retardance near the perimeter of the 1-meter lens, and the necessity to introduce a five-parameter Müller matrix to describe its polarization properties. In future publications [1], we will further explore this data to characterize the imaging properties of this lens for polarization measurements and to search for evidence for significant changes in the polarization properties due to the elevation dependent gravity load on the lens. This may lead to further refinements of the polarization model.

# Appendix A

# Telescope polarization data from May 8, 2005

The full telescope polarization data from May 8, 2005 is shown in figure A.1, and the residual from the telescope model-fitting is shown in figure A.2. Note that the $I$ component is not used in the model-fitting, since only the normalized Stokes components $Q/I$, $U/I$ and $V/I$ are used. Thus for all samples $I = 1$ and does not contribute with any information.

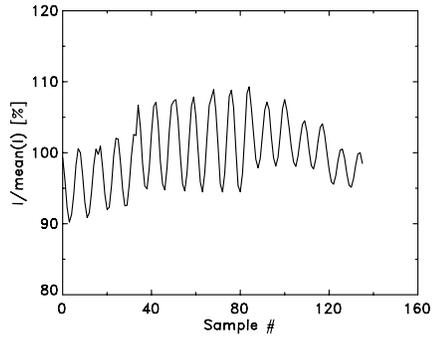
(a) $I$

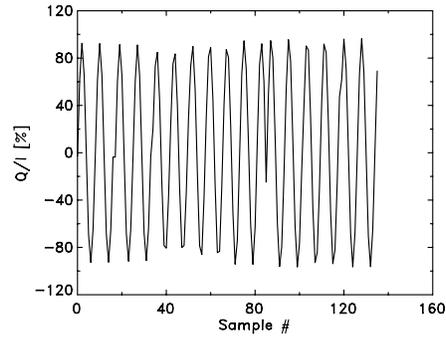
(b) $Q/I$

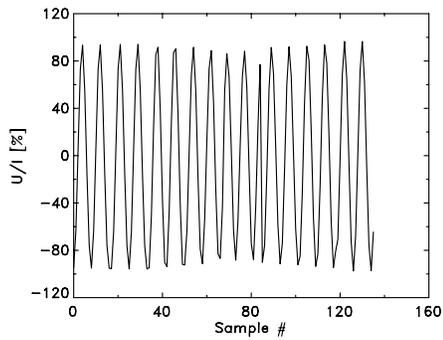
(c) $U/I$

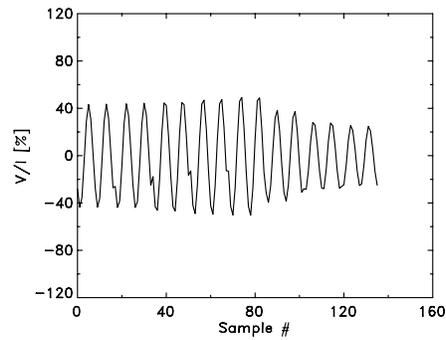
(d) $V/I$

Figure A.1: Complete polarization data from May 8, 2005. Note that Stokes components are normalized with respect to intensity $I$, except $I$ itself which has been normalized with respect to its mean value ($I$ is not used in model-fitting). Discontinuities are due to the plotting against sample number, and not revolution and/or time.

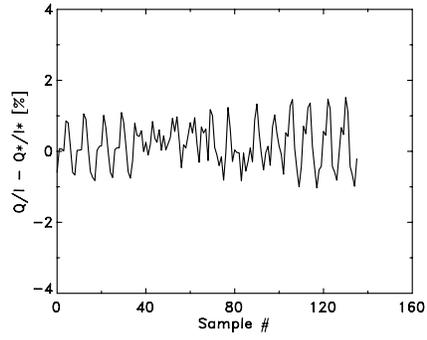

(a) $Q/I$

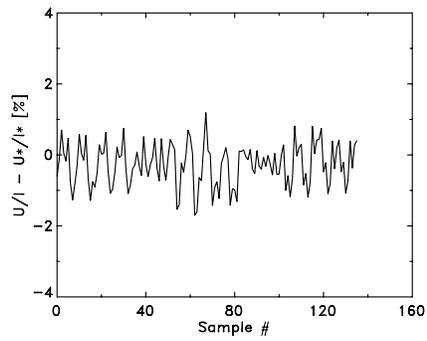

(b) $U/I$

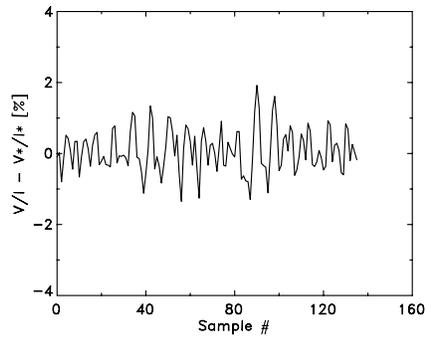

(c) $V/I$

Figure A.2: Model-fitting residual for complete polarization data from May 8, 2005.

# Appendix B

# Stokes vector and Müller matrix formalism

## B.1 Stokes vector

A beam of light can be polarized, i.e. the direction of the oscillation of the electric field is not completely random over time. To describe the polarization, one can use the Stokes vector. The Stokes vector $S$, is composed by four scalars:

$$\boldsymbol{S} = \begin{pmatrix} I & Q & U & V \end{pmatrix}^T \tag{B.1}$$

which are directly related to experimental measuring. Consider four filters, which have transmission dependent upon polarization. The first one, F1, is a neutral filter which transmits half the intensity, independently of polarization. The second one, F2, is a perfect linear polarizer, which allows only the electric field component along a certain direction (usually called positive $Q$ direction). F3 is also a linear polarizer, which is rotated 45° rom the direction of F2 counterclockwise as seen from the observer receiving the light. F4 is a filter that only allows right-circularly polarized light to pass through. If $I_{\text{Fn}}$ is the measured intensity using filter Fn, one can define each Stokes component by the relations

$$I = 2\left(I_{\text{F1}}\right) \tag{B.2}$$
$$Q = 2\left(I_{\text{F2}} - I_{\text{F1}}\right) \tag{B.3}$$
$$U = 2\left(I_{\text{F3}} - I_{\text{F1}}\right) \tag{B.4}$$
$$V = 2\left(I_{\text{F4}} - I_{\text{F1}}\right) \tag{B.5}$$

The $I$ component is simply the absolute intensity. $Q$ is the degree of linearly polarized light along the positive $Q$ direction. $U$ is also the degree of linearly polarized light, but along the direction 45° from positive $Q$. The $V$ component is the degree of right-circularly polarized light. The $Q$, $U$ and $V$ components can be both positive and negative, but the absolute value is always less than or equal to I.

The normalized Stokes vector, which is sometimes more convenient to use, is simply

$$\boldsymbol{S}_\text{n} = \begin{pmatrix} 1 & \frac{Q}{I} & \frac{U}{I} & \frac{V}{I} \end{pmatrix}^T \tag{B.6}$$

The degree of polarization $P$ is defined as

$$P = \left( \frac{Q^2 + U^2 + V^2}{I^2} \right)^{1/2} \tag{B.7}$$

and it always holds that $P \leq 1$.

## B.2  Müller matrix formalism

Most optical elements affect the polarization of transmitted or reflected light. For in principle all elements, the relation between the components of the Stokes vector of the incoming light and the Stokes vector of the outcoming light is a linear function according to

$$I_\text{out} = m_{11}\, I_\text{in} + m_{12}\, Q_\text{in} + m_{13}\, U_\text{in} + m_{14}\, V_\text{in} \tag{B.8}$$
$$Q_\text{out} = m_{21}\, I_\text{in} + m_{22}\, Q_\text{in} + m_{23}\, U_\text{in} + m_{24}\, V_\text{in} \tag{B.9}$$
$$U_\text{out} = m_{31}\, I_\text{in} + m_{32}\, Q_\text{in} + m_{33}\, U_\text{in} + m_{34}\, V_\text{in} \tag{B.10}$$
$$V_\text{out} = m_{41}\, I_\text{in} + m_{42}\, Q_\text{in} + m_{43}\, U_\text{in} + m_{44}\, V_\text{in} \tag{B.11}$$
$$\tag{B.12}$$

or equivalently

$$\boldsymbol{S}_\text{out} = \mathbf{M}\boldsymbol{S}_\text{in} \tag{B.13}$$

where $\mathbf{M}$ is the so-called Müller matrix of the optical element. The Müller matrix is specific for a certain coordinate frame (positive $Q$) for the incoming Stokes vector and outcoming Stokes vector. Thus if the element is rotated one must make a coordinate transformation first for the incoming vector (into the optical element's coordinate frame) and then for the outcoming vector (back to the original coordinate frame). Thus the relation is now

$$\boldsymbol{S}_\text{out} = \mathbf{R}(-\alpha)\, \mathbf{M}\, \mathbf{R}(\alpha)\, \boldsymbol{S}_\text{in} = \text{Rot}(\mathbf{M}, \alpha)\, \boldsymbol{S}_\text{in} \tag{B.14}$$

where $\mathbf{R}(\alpha)$ is the rotational transformation into a coordinate frame at the angle $\alpha$ from the original, and with value

$$\mathbf{R}(\alpha) = \begin{pmatrix} 1 & 0 & 0 & 0 \\ 0 & \cos 2\alpha & \sin 2\alpha & 0 \\ 0 & -\sin 2\alpha & \cos 2\alpha & 0 \\ 0 & 0 & 0 & 1 \end{pmatrix} \tag{B.15}$$

## B.3 Common Müller matrices

### B.3.1 Partial linear polarizer

$$\mathbf{M} = 0.5 \begin{pmatrix} k_1^2 + k_2^2 & k_1^2 - k_2^2 & 0 & 0 \\ k_1^2 - k_2^2 & k_1^2 + k_2^2 & 0 & 0 \\ 0 & 0 & 2\,k_1\,k_2 & 0 \\ 0 & 0 & 0 & 2\,k_1\,k_2 \end{pmatrix} \quad \text{(B.16)}$$

where $k_1$ and $k_2$ is the electrical field attenuation along positive $Q$ and negative $Q$ respectively. If common transmittance is ignored, the matrix is

$$\mathbf{M} = \begin{pmatrix} 1 + K_{\text{LP}} & 1 - K_{\text{LP}} & 0 & 0 \\ 1 - K_{\text{LP}} & 1 + K_{\text{LP}} & 0 & 0 \\ 0 & 0 & 2\sqrt{K_{\text{LP}}} & 0 \\ 0 & 0 & 0 & 2\sqrt{K_{\text{LP}}} \end{pmatrix} \quad \text{(B.17)}$$

where $K_{\text{LP}} = (k_1/k_2)^2$ is the extinction ratio of the polarizer. For convenience one can consider all polarizers having $k_2 \ll k_1$ so that $K_{\text{LP}} \ll 1$ ($10^{-5}$ for a high-precision polarizer). Sometimes instead the contrast $C$ is given, which is

$$C = \frac{1}{K_{\text{LP}}} \quad \text{(B.18)}$$

### B.3.2 Linear retarder

$$\mathbf{M} = \begin{pmatrix} 1 & 0 & 0 & 0 \\ 0 & 1 & 0 & 0 \\ 0 & 0 & \cos\delta & \sin\delta \\ 0 & 0 & -\sin\delta & \cos\delta \end{pmatrix} \quad \text{(B.19)}$$

where $\delta$ is the phase retardance. The fast axis is along positive $Q$. Rotated to an arbitrary angle $\alpha$ the Müller matrix is

$$\widetilde{\mathbf{M}} = \begin{pmatrix} 1 & 0 & 0 & 0 \\ 0 & \cos^2 2\alpha + \sin^2 2\alpha \cos\delta & \sin 2\alpha \cos 2\alpha\,(1-\cos\delta) & -\sin 2\alpha \sin\delta \\ 0 & \sin 2\alpha \cos 2\alpha\,(1-\cos\delta) & \sin^2 2\alpha + \cos^2 2\alpha \cos\delta & \cos 2\alpha \sin\delta \\ 0 & \sin 2\alpha \sin\delta & -\cos 2\alpha \sin\delta & \cos\delta \end{pmatrix} \quad \text{(B.20)}$$

### B.3.3 Linear retarder with dichroism

$$\mathbf{M} = \begin{pmatrix} 1 & b & 0 & 0 \\ b & 1 & 0 & 0 \\ 0 & 0 & (1-b^2)\cos\delta & (1-b^2)\sin\delta \\ 0 & 0 & -(1-b^2)\sin\delta & (1-b^2)\cos\delta \end{pmatrix} \quad \text{(B.21)}$$

where

$$b = \frac{r_x - r_y}{2\,(r_x + r_y)} \quad \text{(B.22)}$$

and $\delta$ is the phase retardance, and $r_x$ and $r_y$ is transmittance along positive and negative $Q$ respectively. Common transmittance is ignored. The fast axis is along positive $Q$.

### B.3.4 Free metallic mirror

$$\mathbf{M} = \begin{pmatrix} 1+R & 1-R & 0 & 0 \\ 1-R & 1+R & 0 & 0 \\ 0 & 0 & -2\sqrt{R}\cos\delta & -2\sqrt{R}\sin\delta \\ 0 & 0 & 2\sqrt{R}\sin\delta & -2\sqrt{R}\cos\delta \end{pmatrix} \quad (B.23)$$

Positive $Q$ is perpendicular to the plane of incidence for both incident and reflected beam. Note that both $\rho$ and $\delta$ depend on physical properties, such as the complex refractive index (2 parameters) of the metal and the angle of incidence. However by letting both parameters be free, one gets a more general matrix but with the same number of parameters.

### B.3.5 Zero-degree mirror

For a metallic mirror with a zero-degree angle of incidence, the Müller matrix is

$$\mathbf{M} = \begin{pmatrix} 1 & 0 & 0 & 0 \\ 0 & 1 & 0 & 0 \\ 0 & 0 & -1 & 0 \\ 0 & 0 & 0 & -1 \end{pmatrix} \quad (B.24)$$

Positive $Q$ is arbitrary, *but equal for incident and reflected beam*.

### B.3.6 General birefringent window

This is a general model of a window with arbitrary birefringence, like vacuum telescope entrance windows [1]. The Müller matrix of the entrance window is

$$\mathbf{M} = \begin{pmatrix} 1 & 0 & 0 & 0 \\ 0 & A & B & -C \\ 0 & B & D & E \\ 0 & C & -E & (A+D-1) \end{pmatrix} \quad (B.25)$$

where, for little depolarization, $A, D \approx 1$ and $B, C, E \approx 0$, Positive $Q$ is arbitrary for this Müller matrix.

## B.4 Trains of zero-degree mirrors and rotations

The Müller matrix of a zero-degree mirror and a rotational transformation to an arbitrary coordinate frame is $\mathbf{M}_0$ and $\mathbf{R}(\alpha)$. When several such matrices are multiplied, the result is

$$\mathbf{M}_0\,\mathbf{R}(\alpha) = \begin{pmatrix} 1 & 0 & 0 & 0 \\ 0 & \cos 2\alpha & \sin 2\alpha & 0 \\ 0 & \sin 2\alpha & -\cos 2\alpha & 0 \\ 0 & 0 & 0 & 1 \end{pmatrix} \quad (\text{B.26})$$

$$\mathbf{R}(\beta)\,\mathbf{M}_0\,\mathbf{R}(\alpha) = \begin{pmatrix} 1 & 0 & 0 & 0 \\ 0 & \cos 2(\alpha-\beta) & \sin 2(\alpha-\beta) & 0 \\ 0 & \sin 2(\alpha-\beta) & -\cos 2(\alpha-\beta) & 0 \\ 0 & 0 & 0 & 1 \end{pmatrix} = \mathbf{M}_0\,\mathbf{R}(\alpha-\beta)$$
(B.27)

$$\mathbf{M}_0\,\mathbf{R}(\beta)\,\mathbf{M}_0\,\mathbf{R}(\alpha) = \mathbf{M}_0\,\mathbf{M}_0\,\mathbf{R}(\alpha-\beta) = \mathbf{R}(\alpha-\beta) \quad (\text{B.28})$$

$$\mathbf{R}(\gamma)\,\mathbf{M}_0\,\mathbf{R}(\beta)\,\mathbf{M}_0\,\mathbf{R}(\alpha) = \mathbf{R}(\gamma)\,\mathbf{R}(\alpha-\beta) \quad (\text{B.29})$$

which covers all cases, even those with only rotations or only reflections. Thus any train of zero-degree mirrors and rotations will be one of two forms: $\mathbf{M}_0\,\mathbf{R}(\alpha)$ for an odd number of mirrors, and $\mathbf{R}(\alpha)$ for an even number of mirrors.

# Appendix C

# Polarimetry

## C.1 Definitions

### C.1.1 Modulation and demodulation matrix

The modulation matrix $P$ describes the relation between modulated intensities $I$ and incident Stokes vector $S$ by

$$\boldsymbol{I} = \mathbf{P}\,\boldsymbol{S} \tag{C.1}$$

The demodulation matrix $\mathbf{D}$ is simply

$$\mathbf{D} = \mathbf{P}^{-1} \tag{C.2}$$

if $\mathbf{P}$ is non-singular. If $\mathbf{P}$ is overderdetermined, then usually least-square inversion is assumed.

### C.1.2 Modulation efficiencies

The modulation efficencies are defined as

$$\epsilon_k = (N \sum_{i=1}^{i=N} D_{k,i}^2)^{-1/2}, \quad k = 1, 2, 3, 4 \tag{C.3}$$

where $\mathbf{D}$ is the demodulation matrix, defined as the inverse of the modulation matrix[1].

---

[1] In practical cases, the *measured* modulation matrix.

# Appendix D

# List of ANA scripts

## D.1 Scripts

**lcvr_cal.ana** Calculates response of an LCVR from two `lccal`-files (and dark file) containing measured intensity for crossed and parallel polarizers (see Section 2.2) . Creates one fz-file containing retardance and voltage, and one PostScript plot.

**lcvr_set.ana** Estimate LCVR modulation voltages for a given target modulation matrix. Does not save anything on disk.

**pol_matrix.ana** Calculate actual modulation matrix, and many other things, from a `matrixcal`-file and a polarizer-only observation file (see Section D.1. Results are stored in polarimeter calibration directory.

**lp_obs.ana** Creates a polarizer-only observation file, from a number of polarimetric observations (images) made with the calibrational linear polarizer in place (but *not* the calibrational quarter-wave plate). The output file is simply the average masked and dark calibrated intensity for a user-defined area in the images, and is stored in the polarimeter calibration directory. The file is needed by the script pol_matrix.ana, for estimating the retardance of the calibrational quarter-wave plate.

**lp_obs.cfg** Configuration file for lp_obs.ana, and has to be edited by user.

**telescope_img_demod.ana** Script for polarimetric subframe averaging and image demodulation for testing purposes (with telescope compensation). It also reads information from a turret log file to output the telescope coordinates at the time of each polarimetric observation. It can be used for processing of telescope polarization data (subframe averaging), and simple polarimetric demodulation. Input images are assumed raw, so dark frames and flat fields are also required as input.

**telescope_img_demod.cfg** Configuration file for the script `telescope_img_demod.ana`.

**polar_demod.ana** Main script for polarimetric image demodulation of corrected images (flat fielded and/or MFBD processed). The file `polar_demod.cfg` contains all information, like which images to demodulate, which processing should be made, etc. It can do destretching, telescope compensation, and other things. It reads both FITS and FZ images. For FITS images the time of observations must be defined as the "TIME-OBS" parameter in the file header, and for FZ images the third token in the file header. The turret log file is needed for telescope compensation.

**polar_demod.cfg** Configuration file for `polar_demod.ana`.

**demod_matrixcal.ana** Quick demodulation of a `matrixcal`-file, using a given demodulation matrix in the polarimeter calibration directory. Result is written to the file `demod_matrixcal_Mstokes.fz`.

**extract_iraw.ana** Extraction (de-interleaving) of modulation intensities I1–I4 and IV1–IV2 from a `matrixcal`-file. Result is written to `extract_iraw_Miraw.fz` and `extract_iraw_Mivraw.fz`.

**fit_qw_full.ana** Calculation of calibrational quarter-wave plate retardance and angle offset using the old method with parallel and crossed polarizers (not used any more).

**nlr_table_full.ana** Calculate non-linearity of a CCD-camera from a `qwcal`-file containing intensity for one rotating polarizer between parallel polarizers (not used any more).

**turret_log_position.ana** Conversion of turret log file to ANA matrix (stored on disk).

## D.2 Function definitions

**Pconst_defs.ana** Definitions for polarimetry scripts, with default values, directories, etc.

**Pfunc_findvoltages.ana** Functions for finding modulation voltage for a given retardance of LCVR.

**Pfunc_fit_lp.ana** Model-fit to intensity for one rotating polarizer between fixed parallel polarizers.

**Pfunc_fitlp_nlr.ana** Model-fit to intensity for one rotating polarizer between fixed parallel/crossed polarizers, using a polynomial or square-root model for the camera response.

**Pfunc_fitlp_nlr_single.ana** Model-fit to intensity for one rotating polarizer between fixed parallel polarizers, using a polynomial or square-root model for the camera response.

**Pfunc_fitnlr_model.ana** Fitting of polynomial non-linear response model, using two vectors of intensities with different exposure time (not used any more).

**Pfunc_fitqw_single.ana** Fit of quarter-wave plate parameters, using intensity for rotation of quarter-wave plate between parallel polarizers (not used any more).

**Pfunc_getmodeindex.ana** Generate indices for a certain modulation intensity in a `matrixcal`-file variable.

**Pfunc_optimize_qw_enhanced.ana** Optimization of quarter-wave parameters in the modulation matrix calculation.

**Pfunc_solvemodangles.ana** Calculate modulation retardances for a given modulation.

**Pfunctions.ana** Various short functions.

## D.3  Library

**fit_sst_optimized_generallens_f.ana** Fitting of simplified telescope model to polarization data (using 1-meter polarizer).

**sdata_fit_sst_optimized_generallens_f.ana** Generate polarization data for a set of telescope coordinates and 1-meter polarizer angles for certain telescope model parameters. Format is interleaved. Use `sdata_mat2arr` and `sdata_arr2mat` to convert between matrix and interleaved format.

**sdata_mat2arr_f.ana** Convert a 4-row matrix to a vector containing the matrix columns one after each other.

**sdata_arr2mat_f.ana** Convert a vector to a 4-row matrix where each column is taken from consecutive elements in the vector.

**udata_generallens_f.ana** Like sdata_fit_sst_optimized_generallens, but for measurements without the 1-meter polarizer.

**telmatrix_generallens_f.ana** Return telescope matrix at a certain coordinate and vector of model parameters.

**turret_log_read_f.ana** Reads a turret log file and creates an ANA matrix containing all the information.